\documentclass[aps,pre,amsmath,amssymb,reprint,numeric,fleqn]{revtex4-1}
\usepackage[english]{babel}
\usepackage{float}
\usepackage{dcolumn}
\usepackage[applemac]{inputenc}
\usepackage{color}
\usepackage{bm}
\usepackage[normalem]{ulem}
\usepackage{multirow}
\usepackage{lipsum}
\usepackage{graphicx}
\usepackage{grffile}
\usepackage{subfigure}

\begin{document}

\title{Fluctuation-Dissipation Relations in the imbalanced Wilson-Cowan model}

\author{Manoj Kumar Nandi$^1$, Antonio de Candia$^{2,3}$, Alessandro Sarracino$^{1,4}$, Hans J. Herrmann$^{5,6}$, Lucilla de Arcangelis$^7$}
\email{lucilla.dearcangelis@unicampania.it}

\affiliation{$^1$ Department of Engineering, University of Campania ``Luigi Vanvitelli'' 81031 Aversa (Caserta), Italy\\
$^2$ Dipartimento di Fisica ``E. Pancini'', Universit\`{a} di Napoli Federico II, Napoli, Italy\\
$^3$ Istituto Nazionale di Fisica Nucleare (INFN), Sezione di Napoli, Gruppo collegato di Salerno, Fisciano, Italy\\
$^4$ Institute for Complex Systems–CNR, Piazzale Aldo Moro 2, 00185, Rome, Italy\\
  $^5$ PMMH, ESPCI, 7 quai St. Bernard, Paris 75005, France\\
$^6$ Departamento de Fisica, Universidade Federal do Cear\'{a}, 60451-970, Fortaleza, Cear\'{a}, Brazil\\
$^7$ Department of Mathematics \& Physics, University of Campania "Luigi Vanvitelli"
Viale Lincoln, 5, 81100 Caserta,  Italy}

\begin{abstract}
 The relation between spontaneous and stimulated brain activity is a
 fundamental question in neuroscience, which has received wide
 attention in experimental studies.
 Recently, it has been suggested
 that the evoked response to external stimuli can be predicted from
 temporal correlations of spontaneous activity. Previous theoretical results,
 confirmed by the comparison with MEG data for human brains, were
 obtained for the Wilson-Cowan model in the condition of balance of
 excitation and inhibition, signature of a healthy brain. Here we
 extend previous studies to imbalanced conditions, by examining a
 region of parameter space around the balanced fixed point. Analytical
 results are compared to numerical simulations of Wilson-Cowan
 networks. We evidence that in imbalance conditions the functional
 form of the time correlation and response functions can show several behaviors,
 exhibiting also an oscillating regime caused by the emergence of
 complex eigenvalues. The analytical predictions are fully in
 agreement with numerical simulations, validating the role of
 cross-correlations in the response function. Furthermore, we identify
 the leading role of inhibitory neurons in controlling the overall
 activity of the system, tuning the level of excitability and
 imbalance.
\end{abstract}

\maketitle
\section{Introduction}
The brain is a complex system whose properties emerge from the
structured interaction between its fundamental constituents, the
neurons. Beyond the stimulated response
by external perturbations, neurons present a background rest activity,
as observed both in vivo and in vitro
experiments~\cite{petermann2009spontaneous,mazzoni,deco2012ongoing,deco2011emerging}.
A natural question arises therefore about the relation between such
two kinds of activity, spontaneous and stimulated. In particular,
clarifying how the brain's response to external stimuli can depend on
the ongoing rest activity could shed light on the main mechanisms
ruling the observed large variability in the
dynamics~\cite{arieli1996dynamics}.  An even more ambitious goal would
be the prediction of the brain response to an external stimulus from
the observation of the unperturbed rest
signal~\cite{papo2014functional,sarracino2020predicting}. 
This
possibility is suggested by statistical mechanics and stochastic
processes theory, whose framework allows one to derive the so-called
Fluctuation-Dissipation Relations
(FDRs)~\cite{marconi2008fluctuation}. They express the
response function of a given variable to an external field in terms of
unperturbed correlation functions of appropriate observables. In recent
years, such relations have been extended beyond the realm of
equilibrium systems, to a very general class of models that exhibit 
non-equilibrium dynamics~\cite{puglisi2017temperature}. 
The problem of forecasting the behavior of a system, and in particular
its response to perturbations, from the study of the past history is a very general
issue and has been addressed in many different physical contexts~\cite{marconi2008fluctuation}. In biological systems, however, there are only
a few cases where such kind of approach has been attempted quantitatively. We can mention the study of evolution in bacteria reported in \cite{sato2003relation}, and the application to the heart rate response \cite{chen2013prediction}.
More recently, in the context of brain dynamics, this framework has been applied  to experimental MEG data of human
brain activity~\cite{sarracino2020predicting} and to stochastic models
for the dynamics of a single neuron~\cite{Lindner}.  In
particular, in~\cite{sarracino2020predicting}, in order to obtain an
explicit form of the FDR to apply to data, the authors considered the
celebrated Wilson-Cowan model for excitatory and inhibitory neuron
populations~\cite{wilson,cowan2016wilson}. In the
linearized version of the model, exact expressions could be derived
for response and correlation functions that were fitted to MEG
data. The main result of this study confirmed that a
prediction on the decay of the response function of brain activity
from the observation of spontaneous fluctuations can be obtained, with
good qualitative agreement between theory and experiments.
Here we extend the analysis to a more general model, that also allows one to consider imbalanced neuron populations,
and explore in more detail the different behaviors that can take place in the parameter space.

The Wilson-Cowan model considers a network of two populations of neurons,
excitatory ($E$) and inhibitory ($I$), that are coupled via four
coefficients that represent the synaptic weights between $E$ and $I$
neurons ($w_{EE}$, $w_{II}$, $w_{IE}$ and $w_{EI}$). A neuron is
activated by an input current, which takes into account the
interaction with other neurons and an external field. The model can be
studied at different levels of coarse-graining, from the microscopic
dynamics of the $N$ single neurons, to the large scale description
which reduces to only two variables, namely the fractions of active
excitatory and inhibitory populations. In the large system size limit, one
can derive two deterministic equations ruling the fixed points of the
population dynamics and two coupled Langevin equations describing
their fluctuations. In previous
papers~\cite{Benayoun2010,plos}, this model was studied in
the particular case of synaptic coefficients that only depend on the
presynaptic neuron, namely $w_{EI}=w_{II}$ and $w_{IE}=w_{EE}$, for
both fully connected and sparse networks. The analysis in a 2D geometry is
reported in~\cite{apicella2022power}.  For this choice of synaptic
weights, the coupling matrix between excitatory and inhibitory
populations takes a triangular form, implying an activity correlation
function with a (double) exponential decay. Moreover, in~\cite{plos}
it was shown that a bona fide critical point can be identified for a
specific value of the parameter $w_0=w_{EE}-w_{II}$, characterized by
a diverging characteristic time in the correlation function and by a
power-law scaling of the activity avalanche distribution.

In this paper, we study the fixed points and the correlation and
response functions in the Wilson-Cowan model in the more general case
where the above constraint on the synaptic strengths is released and
considering different ratios of $E$ and $I$ neurons. This allows us to
address the relevant issue related to the presence of an imbalance
condition in the model. The important role played by the relative
fraction of $I$ neurons in the system behavior has been recently
discussed for integrate and fire models
in~\cite{Raimo2020RoleOI,nandi2022scaling}. 
Previous results for specific values of the parameters (far from the
critical point and in the presence of large external fields) have been
reported in~\cite{Wallace2011, bressloff2010metastable}, where noisy
limit cycles and quasi-cycles in the population dynamics were
observed. See also the models discussed in \cite{copelli2019oscillations, zankoc2017diffusion, piuvezam2023unconventional}. Here we reconstruct the whole phase diagram for the total
activity $\Sigma$ and the imbalance between excitatory and inhibitory
activity $\Delta$ (see below for the exact definition) in a region
around the critical point identified in \cite{plos}, in the limit of
vanishing external field.  Our results unveil the existence of an
abrupt, discontinuous change in the phase diagram, separating a region
of finite activity from a region of almost zero activity.  We then
focus on the correlation and response functions, which show a rich
phenomenology depending on the system parameters, featuring damped
oscillations over several time regimes. Experimental
  studies of these quantities have been conducted, for instance in
  neocortical slices, as reported in \cite{wu1999propagating}, or in
  rat somatosensory cortex cultures \cite{plenz1996generation}. Other
  results can be found for correlations of alpha oscillations in
  ~\cite{linkenkaer2001long}, or for the activity fluctuations in
  cortical areas of the macaque monkey~\cite{murray2014hierarchy}.
  
We then study the relation between correlation and response functions
via the FDRs. In particular, the
unperturbed state, described by Eq.~(\ref{stocastic_sigma1}) below,
represents the spontaneous activity, while the stimulation is applied
through a small perturbation to the initial condition, as detailed in
Eqs.~(\ref{response_eq1}) and~(\ref{resp_num}). In the case of a
comparison with experimental data, a delicate issue can be represented
by the correct modelling of the applied stimulus, as discussed
in~\cite{sarracino2020predicting}, for instance due to the kind of
the specific stimulation.  Finally, we compare the analytical solution of
the linearized model to extensive numerical simulations of the
microscopic dynamics, and study the convergence towards the analytical
predictions as a function of the system
size~\cite{bressloff2010}. Quite surprisingly, we find that, in some
cases, such a convergence is very slow, requiring a huge number of
neurons in the simulations.

The paper is organized as follows. In Sec.~\ref{sec1} we
  introduce the stochastic Wilson-Cowan model. In Sec.~\ref{corr_resp}
  and \ref{sec_numeric} we summarize the analytical results for
  correlation and response functions and provide details on the
  numerical simulations of the model, respectively.  In
  Sec.~\ref{fixed} we discuss the fixed points of the dynamical
  equations and in Sec.~\ref{eigen} we comment on the eigenvalues that
  rule the dynamics. Then, in Sec.~\ref{corr} and Sec.~\ref{resp} we
  discuss the different behaviors observed in the model for
  correlation and response functions, respectively. Finally, in
  Sec.~\ref{con} some conclusions are drawn. In the Appendix we
  provide details on the analytical computations.

\section{The stochastic Wilson-Cowan model}
\label{sec1}
The stochastic version of the Wilson-Cowan model
\cite{wilson,ohira,Benayoun2010} describes the coupled dynamics of a
network of two populations, i.e. $N_E$ excitatory and $N_I$ inhibitory
neurons, with $\chi_{E}$ and $\chi_I$ the fractions of excitatory
($N_E/N$) and inhibitory ($N_I/N$) neurons present in the
network. Each neuron $i$ in the model can be in two states, active
$(a_i=1)$, i.e. a neuron firing an action potential or in its
following refractory period, or quiescent $(a_i=0)$, i.e. a neuron at
rest. The dynamics evolves according to a continuous time Markov process. The
transition rate from active to quiescent state ($1\rightarrow 0$) is
$\alpha$ for all the neurons, while the rate from quiescent to active
state $(0\rightarrow 1)$ depends on an activation function
$f(S_i)$. Here $S_i$ is the total synaptic input of the $i$-th neuron,
which is given by
\begin{equation}
 S_i=\sum_j w_{ij}a_j+h_i,
\end{equation}
where $w_{ij}$ are the synaptic strengths and $h_i$ is an external small field equal for all neurons ($h_i\equiv h=10^{-6}$). In the present work the activation function is chosen to be
\begin{equation}
 f(S)=
 \begin{cases}
  \beta \tanh(S),  &  S>0,\\
  0,        &  S\le 0.
 \end{cases}
\end{equation}
In this study we set $\alpha=0.1 ms^{-1}$ and $\beta=1 ms^{-1}$~\cite{plos} and we consider full connectivity. The outgoing synaptic weights $w_{ij}$ are defined as $\frac{w_{EE}}{N_E}$ for each excitatory to excitatory neurons, $-\frac{w_{EI}}{N_I}$ for inhibitory to excitatory, $\frac{w_{IE}}{N_E}$ for excitatory to inhibitory and $-\frac{w_{II}}{N_I}$ for inhibitory to inhibitory connections. The input of a neuron, $S_i$, only depends on the type of neuron, namely if the $i$-th neuron is excitatory then $S_i=S_E$ and if it is inhibitory then $S_i=S_I$.
Thus for our model
\begin{eqnarray}
S_E&=&\frac{w_{EE} }{N_E}k-\frac{w_{EI} }{N_I}l+h,\nonumber\\
S_I&=&\frac{w_{IE}}{N_E}k-\frac{w_{II} }{N_I}l+h,
\label{eqinputs}
\end{eqnarray}
where $ k$ and $l$ are the number of active excitatory and inhibitory neurons, respectively, evolving according to the master equation \cite{wilson}. In the Gaussian approximation, we can write
\begin{eqnarray}
 k&=&N_E E+\sqrt{N_E}\xi_E\nonumber\\
 l&=&N_I I+\sqrt{N_I}\xi_I,
 \label{kl}
\end{eqnarray}
where $E$ and $I$ are deterministic terms of the active excitatory and inhibitory population, which scale with the population size, and $\xi_E$ and $\xi_I$ are the stochastic fluctuation terms, which scale with the square root of the population size. 
It is then possible to expand the master equation as a Taylor series in $(\xi_E, \xi_I)$ around the deterministic terms $(E, I)$ and, by retaining the two leading terms
in the system size, one obtains the so-called linear noise approximation which provides two sets of coupled differential equations for $(E, I)$ and $(\xi_E, \xi_I)$ \cite{ohira}. In order to more easily interpret the system behavior, it is convenient to introduce the new variables $\Sigma=\chi_E E+\chi_I I$ and $\Delta=\chi_E E-\chi_I I$, which represent the total activity of the system and the imbalance in the activity between the excitatory and inhibitory population.
The Wilson-Cowan equations for the deterministic terms (See Appendix) are given by
\begin{eqnarray}
 \frac{d\Sigma}{dt}&=&-\alpha \Sigma +(\chi_E f(S_E)+\chi_I f(S_I))\nonumber\\
&-&\Sigma \frac{(f(S_E)+f(S_I))}{2}-\Delta \frac{(f(S_E)-f(S_I))}{2}\nonumber\\
 \frac{d\Delta}{dt}&=&-\alpha \Delta +(\chi_E f(S_E)-\chi_I f(S_I))\nonumber\\
 &-&\Sigma \frac{(f(S_E)-f(S_I))}{2}-\Delta \frac{(f(S_E)+f(S_I))}{2},
 \label{wc_deterministic}
\end{eqnarray}
where the input currents are written as 
\begin{eqnarray}
S_E&=&w_{EE}E-w_{EI}I+h\nonumber\\
S_I&=&w_{IE}E-w_{II}I+h,
\label{input_strength}
\end{eqnarray}
    and the dynamical equations for $E$ and $I$ are 
    \begin{eqnarray}
 \frac{dE}{dt}&=&-\alpha E+(1-E) f(S_{E})\nonumber\\
  \frac{dI}{dt}&=&-\alpha I+(1-I) f(S_{I}).
  \label{determin22}
    \end{eqnarray}
    
Conversely, the linearized Langevin equations for the fluctuating variables can be expressed as \cite{bressloff2010, Benayoun2010, Wallace2011, plos} (see also the Appendix)
\begin{equation}
 \frac{d}{dt}\begin{pmatrix}
  \xi_\Sigma \\
  \xi_\Delta
\end{pmatrix}={\bf {A}}\begin{pmatrix}
  \xi_\Sigma \\
  \xi_\Delta
\end{pmatrix}+{\bf D}\begin{pmatrix}
  \eta_\Sigma \\
  \eta_\Delta
\end{pmatrix},
\label{stocastic_sigma1}
\end{equation}
where the coefficients of {\bf A} are calculated using
the stationary solutions of Eq.(\ref{wc_deterministic}) and {\bf D} is the amplitude matrix of
the independent white-noise variables $\eta_\Sigma$ and $\eta_\Delta$,
which satisfy $\langle\eta_i(t)\rangle=0$ and
$\langle\eta_i(t)\eta_j(t')\rangle=\delta_{ij}\delta(t-t')$. A
similar two-variable model has been also introduced in the context of
moderately dense fluids to study the dynamics of a massive
tracer~\cite{crisanti2012nonequilibrium,sarracino2010irreversible}. 
 The details of the calculations and the coefficients of
the matrix {\bf A} and {\bf D} are given in the Appendix.
Notice that so far no hypothesis is made on the value of the $w$'s.

  \begin{figure*}[ht]
	\centering
	\subfigure{
		\includegraphics[width=0.35\textwidth]{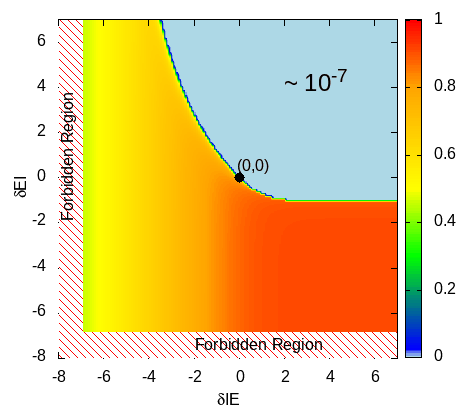}}
	\subfigure{
		\includegraphics[width=0.35\textwidth]{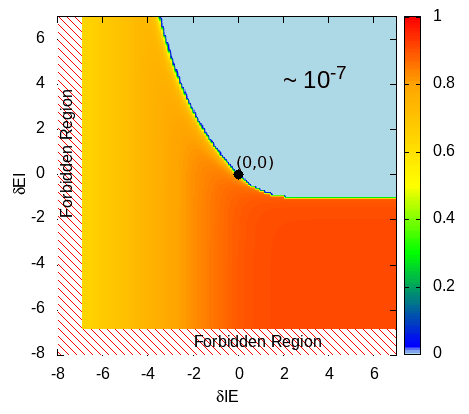}}
	\subfigure{
		\includegraphics[width=0.35\textwidth]{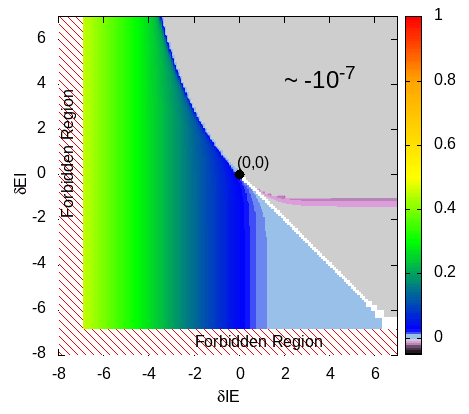}}
	\subfigure{
		\includegraphics[width=0.35\textwidth]{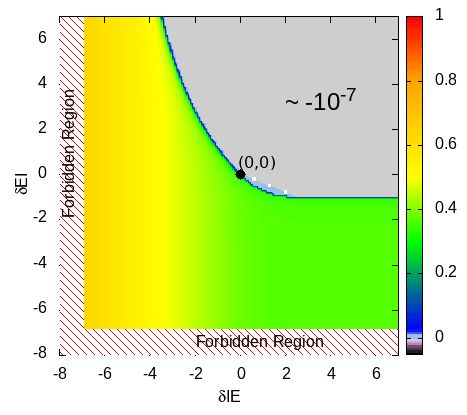}}
	\caption{Values of $\Sigma_0$ (top) and $\Delta_0$ (bottom), solutions of Eq.(\ref{wc_deterministic}), as function of the parameters measuring the variation in the synaptic strengths. The left column refers to $\chi_E=\chi_I=50\%$ and the right column is for $\chi_E=70\%$ and $\chi_I=30\%$. The parameter values for all the calculations are: $w_{EE}=6.95$, $w_{II}=6.85$, $h=10^{-6}$, $\alpha=0.1$, $w_{EI}=w_{II}+\delta {EI}$ and $w_{IE}=w_{EE}+\delta{IE}$. The forbidden regions correspond to the parameter range where $w_{IE}$ or $w_{EI}$ become negative.}
	\label{fig1}
\end{figure*}

\subsection{Correlation and response functions}
\label{corr_resp}

The correlation matrix for the Eq. (\ref{stocastic_sigma1}) in the stationary state can be written as~\cite{crisanti2012nonequilibrium}
\begin{equation}
 C_{ij}(t)\equiv\langle\xi_i(t)\xi_j(0)\rangle=(e^{ {\bf A}t}\sigma)_{ij},
 \label{corr1}
\end{equation}
where angular brackets $\langle\cdots\rangle$ denote average over noise, $(i,j)=(\Sigma, \Delta)$, {\bf $\sigma$} is the covariance matrix and $t\geq 0$.
The correlation functions are 
\begin{eqnarray}
 C_{\Sigma\Sigma}(t)&=&\frac{1}{\sqrt{-\Theta}}[((x-\lambda_2)\sigma_{11}+y\sigma_{21})e^{\lambda_1t}\nonumber\\
 &-&((x-\lambda_1)\sigma_{11}+y\sigma_{21})e^{\lambda_2t}],
 \label{cor11}
\end{eqnarray}
\begin{eqnarray}
 C_{\Sigma\Delta}(t)&=&\frac{1}{\sqrt{-\Theta}}[((x-\lambda_2)\sigma_{12}+y\sigma_{22})e^{\lambda_1t}\nonumber\\
 &-&((x-\lambda_1)\sigma_{12}+y\sigma_{22})e^{\lambda_2t}],
 \label{cor12}
\end{eqnarray}
\begin{eqnarray}
 C_{\Delta\Sigma}(t)&=&\frac{1}{\sqrt{-\Theta}}[(z\sigma_{11}-(x-\lambda_1)\sigma_{21})e^{\lambda_1t}\nonumber\\
 &-&(z\sigma_{11}-(x-\lambda_2)\sigma_{21})e^{\lambda_2t}],
 \label{cor21}
\end{eqnarray}
\begin{eqnarray}
 C_{\Delta\Delta}(t)&=&\frac{1}{\sqrt{-\Theta}}[(z\sigma_{12}-(x-\lambda_1)\sigma_{22})e^{\lambda_1t}\nonumber\\
 &-&(z\sigma_{12}-(x-\lambda_2)\sigma_{22})e^{\lambda_2t}],
 \label{cor22}
\end{eqnarray}
where $\lambda_{1,2}$ are the eigenvalues of the matrix {\bf A}, $x,
y, z$ and $w$ its coefficients which are functions of the model
parameters and $\sqrt{-\Theta}=\sqrt{(x-w)^2+4 yz}$ (see
Appendix). The behavior of the correlation functions is a double
exponential decay with characteristic times $\tau_1=1/\lambda_1$ and
$\tau_2=1/\lambda_2$.  In case of complex eigenvalues,
i.e. $\lambda_{1}=a+ ib$, $\lambda_{2}=a- ib$, where $a=(x+w)/2$ and
$ib=\sqrt{-\Theta}/2$, the correlation functions show oscillatory
behavior with frequency $b$
\begin{eqnarray}
 C_{\Sigma\Sigma}(t)&=&\sigma_{11}e^{at}\cos(bt)+\frac{(x-a)\sigma_{11}+y\sigma_{21}}{b}e^{at}\sin(bt)\nonumber\\
  C_{\Sigma\Delta}(t)&=&\sigma_{12}e^{at}\cos(bt)+\frac{(x-a)\sigma_{12}+y\sigma_{22}}{b}e^{at}\sin(bt)\nonumber\\
   C_{\Delta\Sigma}(t)&=&\sigma_{21}e^{at}\cos(bt)+\frac{z\sigma_{11}-(x-a)\sigma_{21}}{b}e^{at}\sin(bt)\nonumber\\
   C_{\Delta\Delta}(t)&=&\sigma_{22}e^{at}\cos(bt)+\frac{z\sigma_{12}-(x-a)\sigma_{22}}{b}e^{at}\sin(bt),\nonumber\\
   \label{im_cor}
\end{eqnarray}
with $a<0$. The linearization around the fixed points is valid only for values of the parameters for which the matrix {\bf A} has stable eigenvalues.

We next evaluate the linear response function of the system to an instantaneous weak perturbation, defined as
\begin{equation}
 R_{ij}(t)\equiv \frac{\overline{\delta \xi_i(t)}}{\delta \xi_j(0)},
 \label{response_eq1}
\end{equation}
where $(i,j)=(\Sigma, \Delta)$. Eq. (\ref{response_eq1}) represents the average response of $\xi_i(t)$ at time $t$ to the applied pulse perturbation on the variable $\xi_j(0)$ at time $t=0$. The symbol $\overline{(\ldots)}$ denotes the nonstationary average over the trajectories. The response matrix of the system can be calculated for $t>0$ as
\begin{equation}
 {\bf R}(t)= e^{{\bf A}t}.
 \label{response_eq2} 
\end{equation}
From Eq.s (\ref{corr1}) and (\ref{response_eq2}) we can obtain the FDRs which connect the linear response to the spontaneous fluctuations \cite{sarracino2020predicting} as
\begin{equation}
 {\bf R}(t)={\bf C}(t)\sigma^{-1}.
\end{equation}
The response functions for the small perturbations expressed in terms of $\Sigma$ and $\Delta$ are therefore
\begin{eqnarray}
 R_{\Sigma\Sigma}(t)&=&(\sigma^{-1})_{11}C_{\Sigma\Sigma}(t)+(\sigma^{-1})_{21}C_{\Sigma\Delta}(t)\nonumber\\
R_{\Sigma\Delta}(t)&=&(\sigma^{-1})_{12}C_{\Sigma\Sigma}(t)+(\sigma^{-1})_{22}C_{\Sigma\Delta}(t)\label{response_eq3}\\
R_{\Delta\Sigma}(t)&=&(\sigma^{-1})_{11}C_{\Delta\Sigma}(t)+(\sigma^{-1})_{21}C_{\Delta\Delta}(t)\nonumber\\
R_{\Delta\Delta}(t)&=&(\sigma^{-1})_{12}C_{\Delta\Sigma}(t)+(\sigma^{-1})_{22}C_{\Delta\Delta}(t),\nonumber
\end{eqnarray}
where the exact form of the covariance matrix $\sigma$ is
given in the Appendix. From Eq. (\ref{response_eq3}), we see that both
the autocorrelations and the cross correlations are required to
calculate the response to a weak instantaneous perturbation.

In Ref.~\cite{sarracino2020predicting}, the model was
  applied to describe MEG data of humain brains of healthy
  subjects. It was shown that correlation and response functions could
  be well fitted by the model, with parameters corresponding to the
  case of the balance condition. In particular, correlation functions
  show double exponential decay, while
  response functions are well described by a sinlge exponential
  decay.

\begin{figure*}[ht]
	\centering
	\subfigure{
		\includegraphics[width=0.35\textwidth]{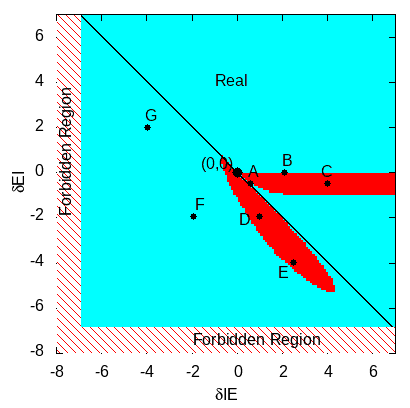}}
	\subfigure{
		\includegraphics[width=0.38\textwidth]{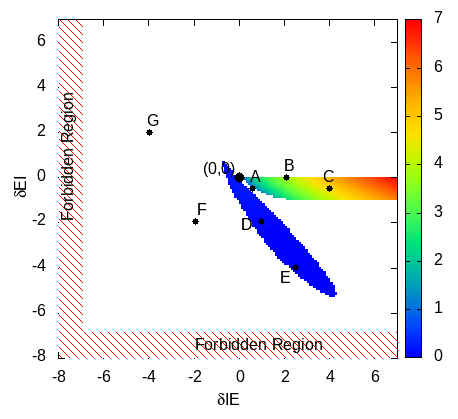}}
	\subfigure{
		\includegraphics[width=0.35\textwidth]{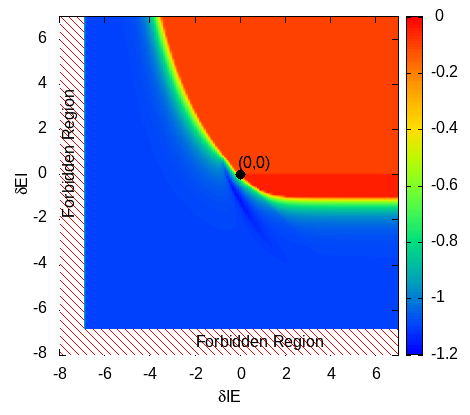}}
		\subfigure{   
		\includegraphics[width=0.35\textwidth]{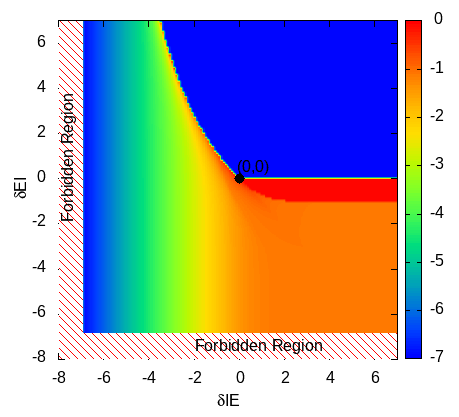}}
               
	\caption{(Top left panel) Eigenvalue phase diagram for different imbalance in synaptic strengths. The cyan color identifies the region with real eigenvalues, whereas the red color indicates the region where eigenvalues are complex. (Top right panel) The values of the imaginary part of the eigenvalues. Letters identify the points in the phase diagram for which correlation and reponse functions have been evaluated. 
	(Bottom left and right panels) The values of the real part of the eigenvalues $\lambda_1$ and $\lambda_2$, respectively. The forbidden regions correspond to the parameter range where $w_{IE}$ or $w_{EI}$ become negative. Eigenvalues are independent of the population size. The numerical parameter values for all the calculations are: $w_{EE}=6.95$, $w_{II}=6.85$, $h=10^{-6}$, $\alpha=0.1$, $w_{EI}=w_{II}+\delta {EI}$ and $w_{IE}=w_{EE}+\delta{IE}$.} 
	\label{eigen1}
\end{figure*}

\subsection{Numerical simulation methods}
\label{sec_numeric}

The continuous time Markov process that describes the dynamics of the system can be efficiently
simulated by the Gillespie algorithm \cite{algoGillespie}, that we describe here for completeness.
The configuration of the system at a given time $t$ is completely determined by the number of
active excitatory neurons $k$, and the number of active inhibitory neurons $l$.
Given $k$ and $l$ at time $t$, we compute the synaptic inputs $S_E$ and $S_I$ from Eq.\ (\ref{eqinputs}), and
then the activation (a) and deactivation (d) rates for excitatory (e) and inhibitory (i) neurons, that are given by
\begin{eqnarray}
r_{de}&=&\alpha k,\nonumber\\
r_{di}&=&\alpha l,\nonumber\\
r_{ae}&=&(N_E-k)f(S_E),\nonumber\\
r_{ai}&=&(N_I-l)f(S_I),
\end{eqnarray}
and the total rate $r_{\text{tot}}=r_{de}+r_{di}+r_{ae}+r_{ai}$.
As the process is Markovian, the time interval to the next event is
extracted from an exponential distribution, $P(\Delta t)=r_{\text{tot}}\exp(-r_{\text{tot}}\Delta t)$,
and the event is selected among $de$, $di$, $ae$ and $ai$ with probability $\frac{r_{de}}{r_{\text{tot}}}$, $\frac{r_{di}}{r_{\text{tot}}}$,
$\frac{r_{ae}}{r_{\text{tot}}}$, $\frac{r_{ai}}{r_{\text{tot}}}$, respectively.
Then the time is incremented by $\Delta t$ and the selected event is performed
increasing by one, or decreasing by one, the number of active neurons $k$ or $l$.

Because in the Gillespie algorithm the time step $\Delta t$ is proportional to $N^{-1}$,
for very large number of neurons the simulation becomes very inefficient. In this case, one can simulate the model using the non-linear Langevin equations 
\begin{eqnarray}
\frac{dk}{dt}&=&-\alpha k+(N_E-k)f(S_E)\nonumber\\
&+&\sqrt{\alpha k+(N_E-k)f(S_E)}\,\eta_E(t),\nonumber\\
\frac{dl}{dt}&=&-\alpha l+(N_I-l)f(S_I)\nonumber\\
&+&\sqrt{\alpha l+(N_I-l)f(S_I)}\,\eta_I(t).
\end{eqnarray}
with a fixed time step. The non-linear equations are equivalent to the full master equation (Gillespie algorithm),
provided that the time step of integration is small enough, and the number of neurons is not too small \cite{plos}.
Here we used a time step of $\Delta t=10^{-3}$ ms. Data for correlation and response functions are averaged over about $10^7$ realizations.

After an appropriate time interval, the process reaches stationarity, so that $k$ and $l$ fluctuate
around their mean values $\bar k=N_E \bar E$ and $\bar l=N_I \bar I$, where $\bar E$ and $\bar I$ are the time average value of
the deterministic components. At stationarity, the fluctuations can be computed by
\begin{align}
\xi_E(t)&=N_E^{-1/2}(k-\bar k),\nonumber\\
\xi_I(t)&=N_I^{-1/2}(l-\bar l),
\end{align}
and
\begin{align}
\xi_\Sigma(t)&=\chi_E\xi_E(t)+\chi_I\xi_I(t)\nonumber\\
\xi_\Delta(t)&=\chi_E\xi_E(t)-\chi_I\xi_I(t).
\end{align}
In this way it is possible to compute the auto-correlations and cross-correlations $C_{\Sigma\Sigma}(t)$, $C_{\Sigma\Delta}(t)$, $C_{\Delta\Sigma}(t)$ and $C_{\Delta\Delta}(t)$ from Eq. (\ref{corr1}).

We next evaluate the response function matrix ${\bf R}(t)$ using the following procedure.
To compute $R_{\Sigma\Sigma}(t)$ and $R_{\Delta\Sigma}(t)$, after the process has reached stationarity,
at a given time $t^\prime$, $k$ and $l$ are perturbed in such a way that
\begin{align}
  \label{resp_num}
\xi_\Sigma(t^\prime)&\rightarrow\xi_\Sigma(t^\prime)+\epsilon,\nonumber
\\
\xi_\Delta(t^\prime)&\rightarrow\xi_\Delta(t^\prime),
\end{align}
where $\epsilon$ is a small quantity. This is performed by
 increasing $k$ by $\frac{\sqrt{N_E}}{2\chi_E}\epsilon$, and $l$ by $\frac{\sqrt{N_I}}{2\chi_I}\epsilon$.
Then we compute
\begin{align}
R_{\Sigma\Sigma}(t)&=\epsilon^{-1}\langle\xi_\Sigma(t^\prime+t)\rangle,\nonumber
\\
R_{\Delta\Sigma}(t)&=\epsilon^{-1}\langle\xi_\Delta(t^\prime+t)\rangle,
\end{align}
where the average $\langle\cdots\rangle$ is done on the realization of the noise, and on different starting configurations at time $t^\prime$.
To compute $R_{\Sigma\Delta}(t)$ and $R_{\Delta\Delta}(t)$ we use a similar procedure, with the difference that we increase $\xi_\Delta(t^\prime)$
instead of $\xi_\Sigma(t^\prime)$. In this case, one has to increase $k$ by $\frac{\sqrt{N_E}}{2\chi_E}\epsilon$ and decrease $l$ by $\frac{\sqrt{N_I}}{2\chi_I}\epsilon$.

\begin{figure*}[ht]
	\centering
	\subfigure{
		\includegraphics[width=0.35\textwidth]{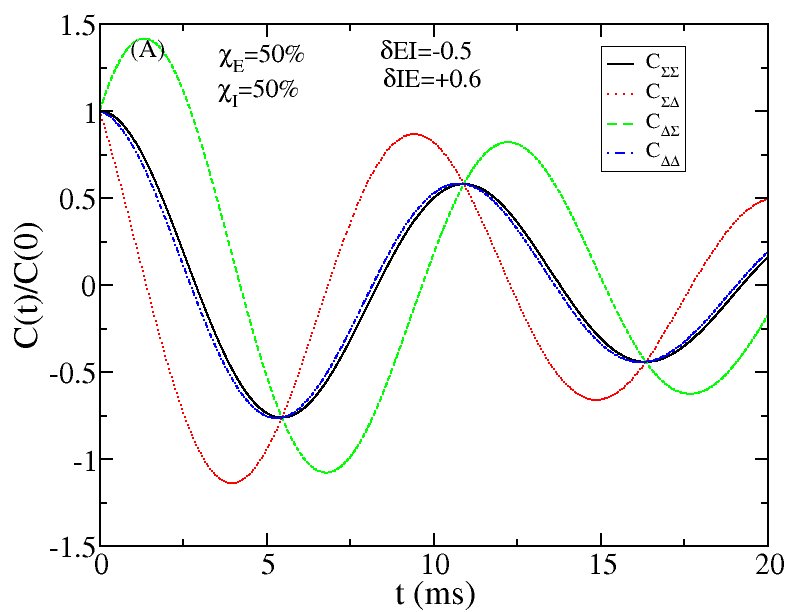}}
	\subfigure{
		\includegraphics[width=0.35\textwidth]{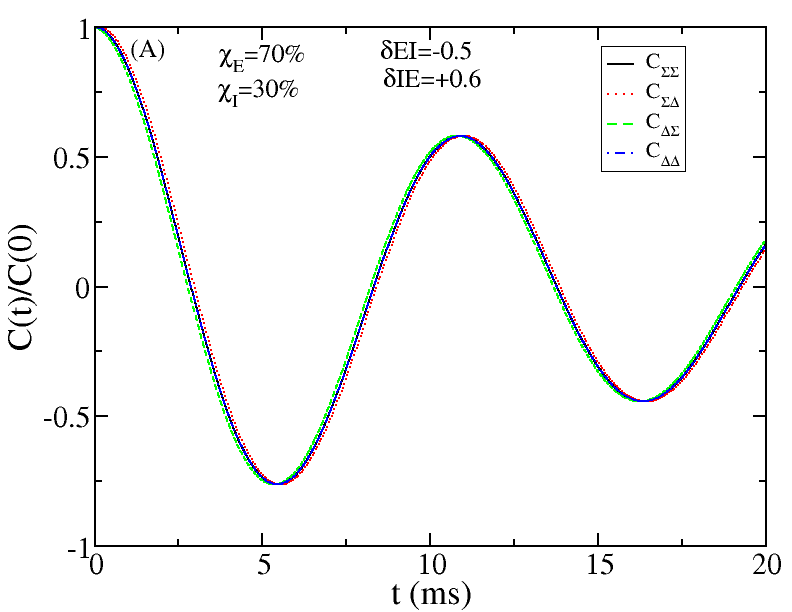}}
	\subfigure{
		\includegraphics[width=0.35\textwidth]{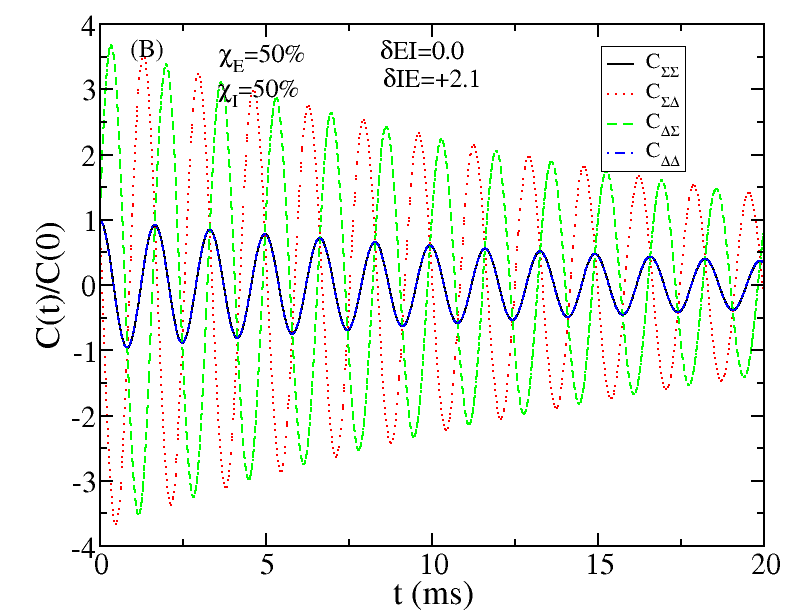}}
	\subfigure{
		\includegraphics[width=0.35\textwidth]{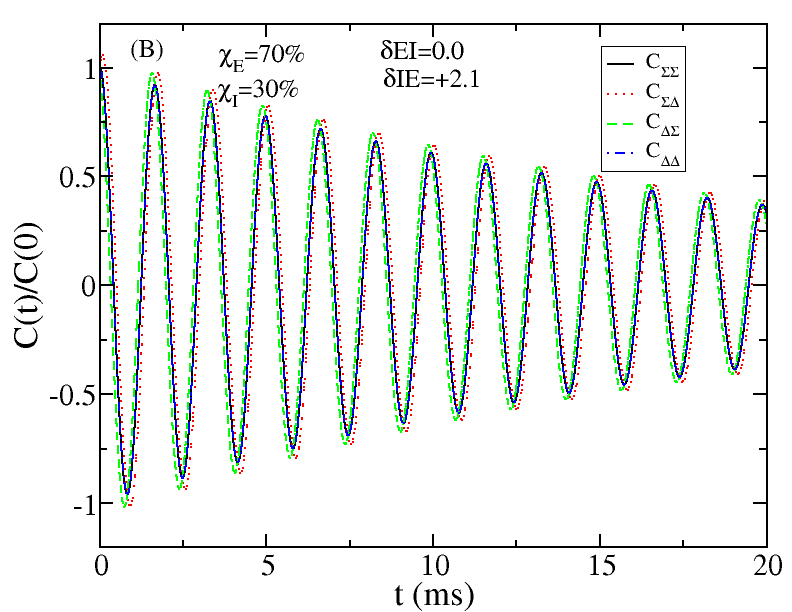}}
	\subfigure{
		\includegraphics[width=0.35\textwidth]{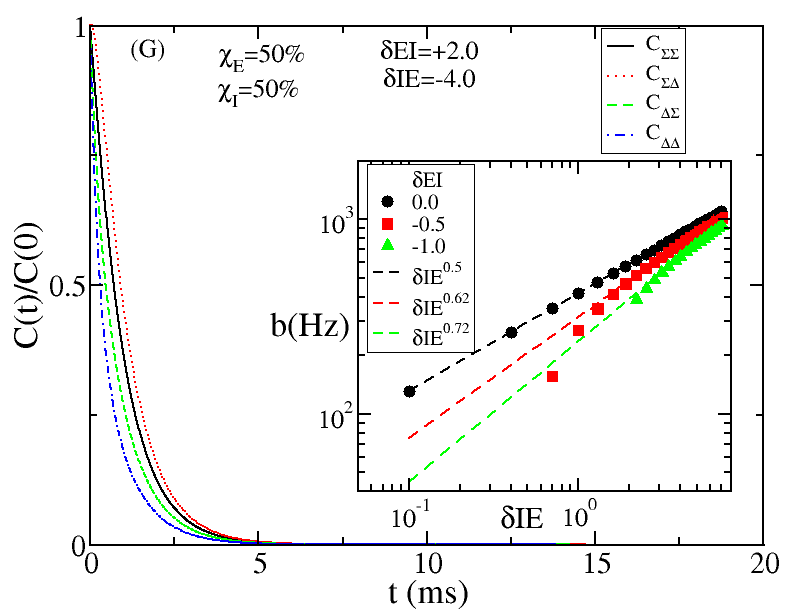}}
	\subfigure{
		\includegraphics[width=0.35\textwidth]{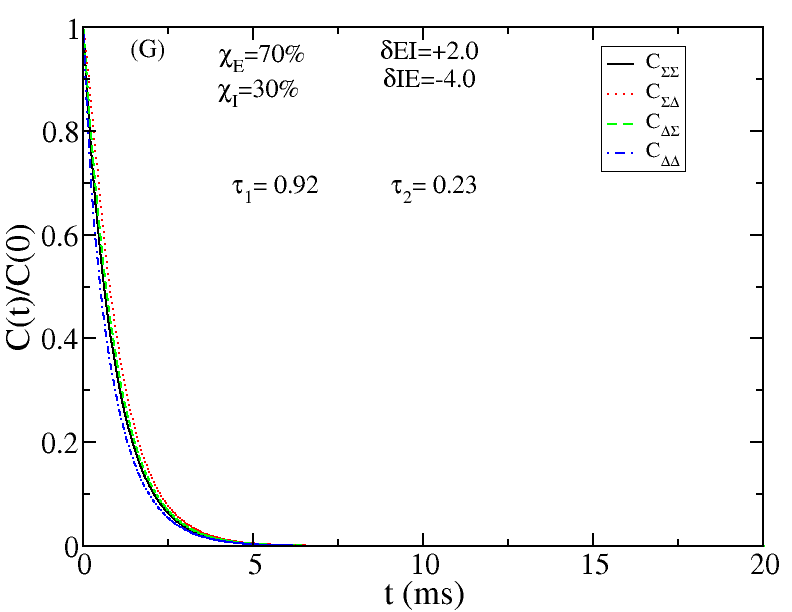}}
	
	\caption{The correlation functions evaluated at the different points A, B, G in the eigenvalue phase diagram Fig.\ref{eigen1}. The left column is for $\chi_E=\chi_I=50\%$ and the right column is for $\chi_E=70\%$ and $\chi_I=30\%$. The inset shows the oscillation frequency $b$ as function of $\delta IE$ for different values of $\delta EI$. The parameter values used for the solution of Eqs.~(\ref{wc_deterministic}) are: $w_{EE}=6.95$, $w_{II}=6.85$, $h=10^{-6}$, $\alpha=0.1$, $w_{EI}=w_{II}+\delta {EI}$ and $w_{IE}=w_{EE}+\delta{IE}$. Time is measured in ms.}
	\label{corr_func_analytic}
\end{figure*}

 \begin{figure*}[ht]
	\centering
	\subfigure{
		\includegraphics[width=0.35\textwidth]{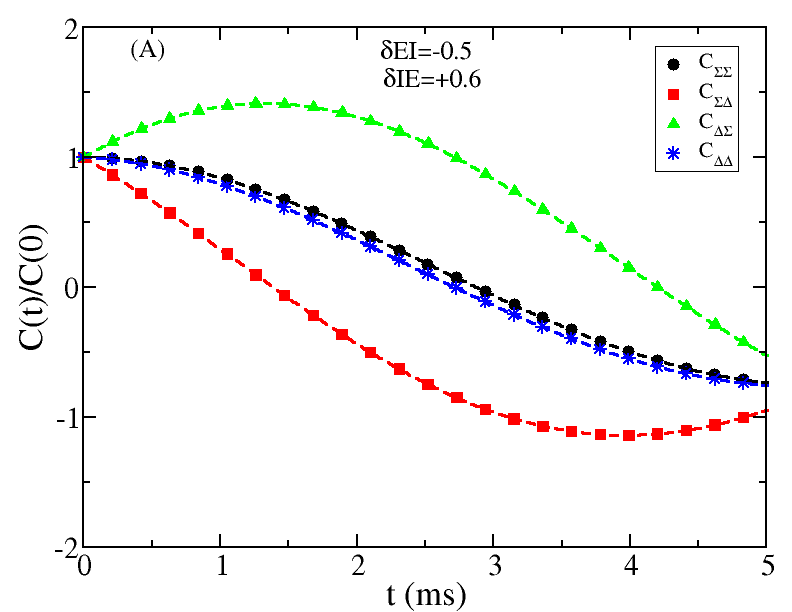}}
	\subfigure{
		\includegraphics[width=0.35\textwidth]{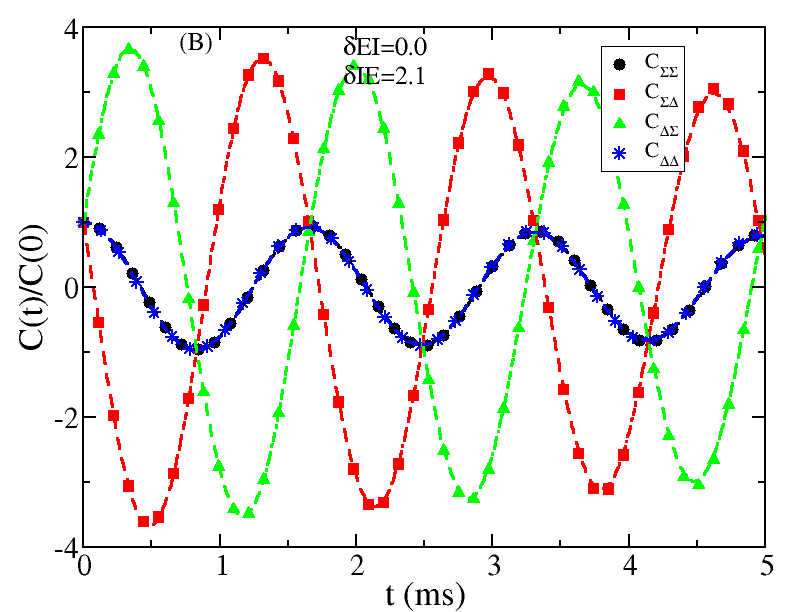}}
	\subfigure{
		\includegraphics[width=0.35\textwidth]{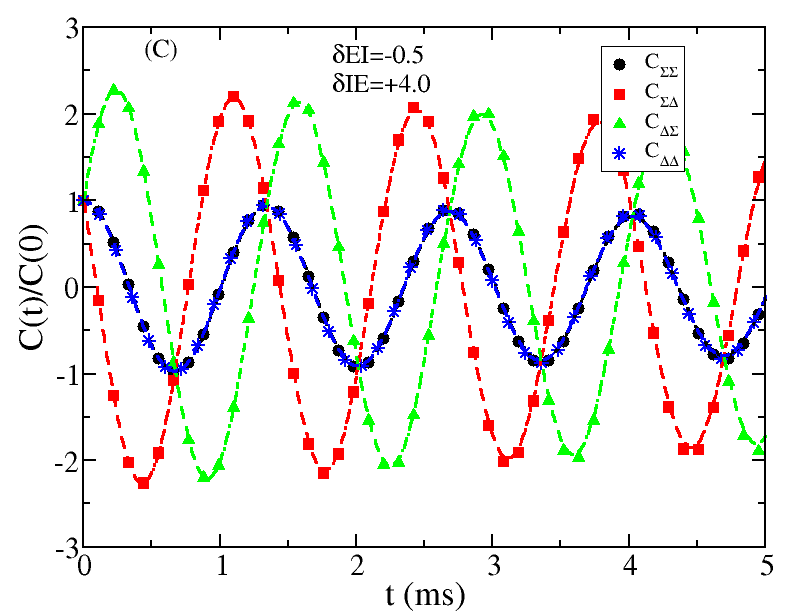}}
	\subfigure{
		\includegraphics[width=0.35\textwidth]{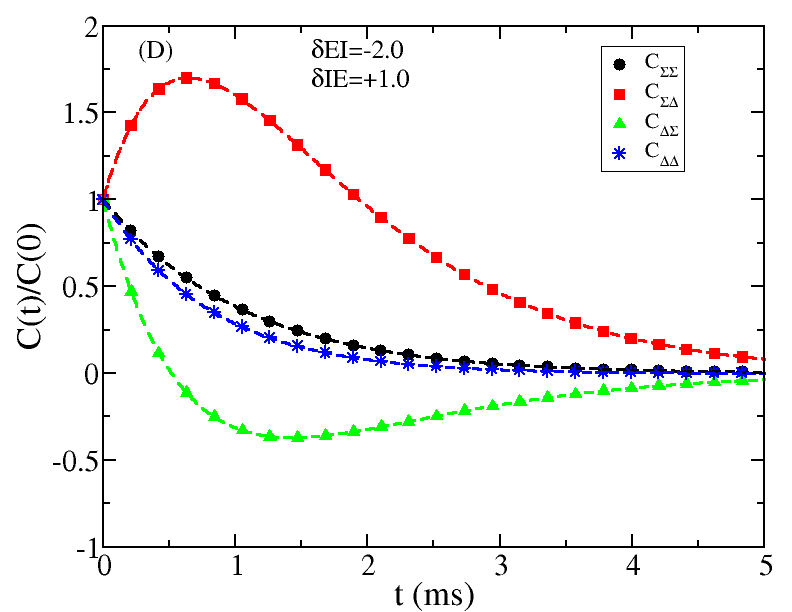}}
	\subfigure{
		\includegraphics[width=0.35\textwidth]{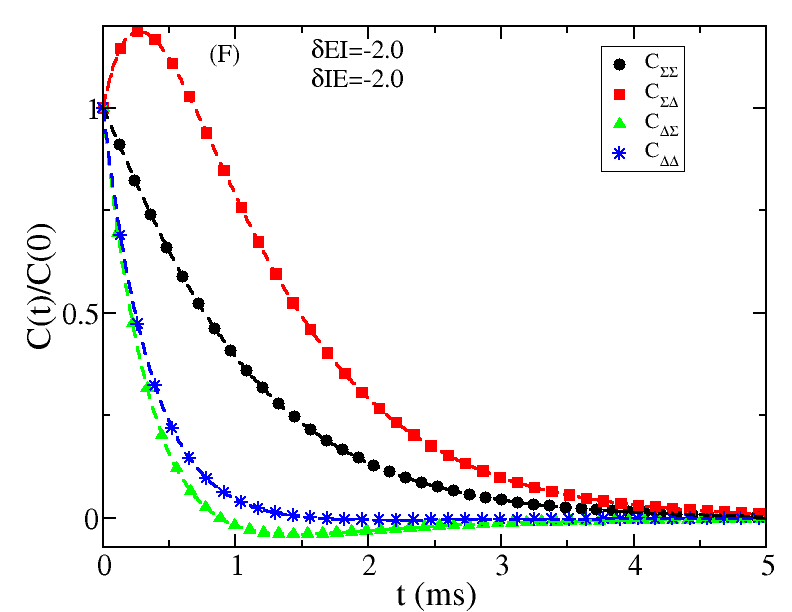}}
	\subfigure{
		\includegraphics[width=0.35\textwidth]{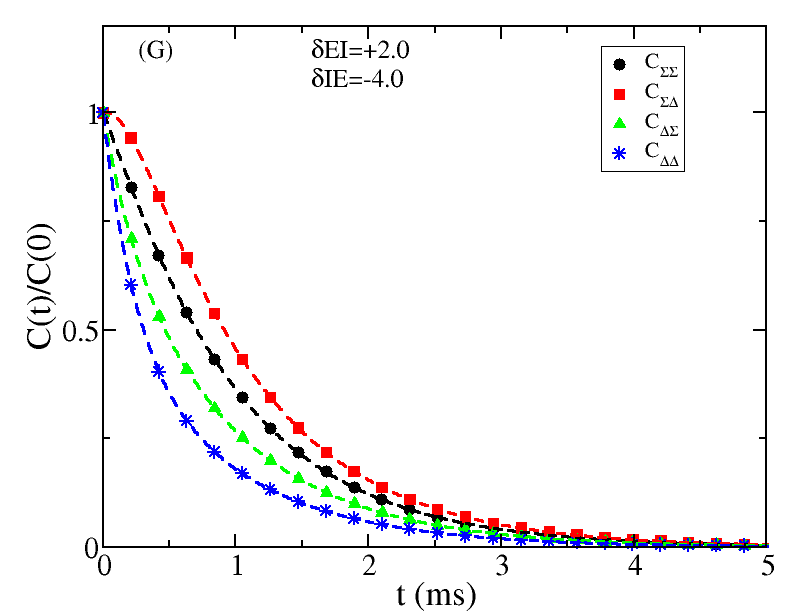}}
	
	\caption{The correlation functions for $\chi_E=\chi_I=50\%$
          evaluated at different points A, B, C, D, F and G in the
          phase diagram Fig.\ref{eigen1}. The symbols are the results of simulations with $N=10^{14}$ neurons,  with parameters  $w_{EE}=6.95$, $w_{II}=6.85$, $h=10^{-6}$, $\alpha=0.1$, $w_{EI}=w_{II}+\delta {EI}$ and $w_{IE}=w_{EE}+\delta{IE}$, whereas dashed lines represent analytical results. Details of simulations are reported in Sec.~\ref{sec_numeric}.}
	\label{corr_compare}
\end{figure*}

 \section{Fixed points}
 \label{fixed}
 
In previous studies, usually synaptic strengths are assumed to solely
depend on the type of presynaptic neuron, namely
$w_{EI}=w_{II}$ and $w_{IE}=w_{EE}$. Following this assumption, the
matrix ${\bf A}$ has an upper triangular form and the fixed point for
imbalance in activity is $\Delta^*=0$. Therefore the previous
condition on synaptic strengths sets the system in a state realizing
the balance of excitatory and inhibitory activity, which leads to the
presence of a critical point at a specific value of
$w_0=w_{EE}-w_{II}=\alpha/\beta=0.1$, where also $\Sigma$ tends to vanish
\cite{plos}.  In the present study, we focus on the behavior of the
system following the removal of such hypothesis, namely by slowly
driving it out of the balance of excitation and inhibition. We start
by fixing the values of the strengths $w_{EE}=6.95$ and $w_{II}=6.85$,
setting the system at criticality. We then define
$w_{EI}=w_{II}+\delta EI $ and $w_{IE}=w_{EE}+\delta IE$, where
$\delta EI$ and $\delta IE$ are the two control parameters tuning the
imbalance condition. Structural inhibition is also tuned by analysing
systems with different fractions of inhibitory neurons,
i.e. $\chi_E=\chi_I=50\%$ and $\chi_E=70\%$, $\chi_I=30\%$. We numerically solve
the deterministic equations (\ref{wc_deterministic}) using Newton's method for fixed point analysis, for different
values of $\delta EI$ and $\delta IE$, in a range corresponding to
positive synaptic connections, to derive the values of
$\Sigma_0$ and $\Delta_0$ at the fixed point (Fig.\ref{fig1}).

The activity $\Sigma_0$ for ($\chi_E=\chi_I=50\%$) and for
($\chi_E=70\%,\chi_I=30\%$) shows similar behaviors. In the regime
where $\delta EI$ is negative, if we change $\delta IE$ from positive
to negative values, the activity $\Sigma_0$ initially shows a plateau
near one and then gradually decreases to 50\% of its initial value. On
the other hand if we change $\delta EI$ keeping $\delta IE$ fixed, the
activity $\Sigma_0$ is almost constant up to certain values of $\delta
EI$ where the activity drops drastically to a very small value ($\sim
10^{-7}$), as shown in Fig.\ref{fig1}, giving rise to a boomerang-like
transition line from a finite to a very small 
activity. For a system with $\chi_E=70\%$ and $\chi_I=30\%$ the
activity $\Sigma_0$ shows a similar behavior, with small
discrepancies with respect to the system with $\chi_E=\chi_I=50\%$ for
very negative $\delta IE$. This behavior can be understood by
considering the different role of the perturbations in the dynamics: A
large variation $\delta IE$ implies that the synaptic connections from
the excitatory to the inhibitory population are stronger than in the
case of balanced activity, leading to an increased activity of the
inhibitory population. Conversely, large $\delta EI$s imply that the
inhibitory population strongly hampers the activity of the excitatory
one. As a consequence, the system activity stems from the interplay
between the relative role of the two populations. The observation that
the excitability of the system strongly increases below the bisector
$\delta IE=-\delta EI$, suggests that the imbalance in excitation is
mostly controlled by the inhibitory population, whose activity cannot
compensate the excitatory one either because of its weak connections
($\delta EI$ is too small) or because they are weakly stimulated by the excitatory population ($\delta IE$ is too small). Interestingly,
in the first quadrant the $\delta$'s are both positive which allows
balance to be achieved within a wide range of parameters.

Conversely, a more clear sensitivity to parameters is observed for the
imbalance $\Delta_0$ at the fixed point. For systems with equal size
populations ($\chi_E=\chi_I=50\%$), if we progressively decrease
$\delta IE$, keeping $\delta EI \leq 0$ fixed, the imbalance
$\Delta_0$ starts from a very small negative value ($\sim -10^{-7}$),
vanishes at the bisector line $\delta IE=-\delta EI$ and gradually
increases to a maximum value of $\sim 0.5$. Conversely, if we change
$\delta EI$, keeping $\delta IE \ge 0$ fixed, $\Delta_0$ appears to be
roughly independent of $\delta EI$ but abruptly drops to a very small
negative value ($\sim -10^{-7}$) for parameter values above the
bisector line. Activity therefore appears imbalanced in favor of
excitation in a wide region of parameters corresponding to large
positive $\Sigma_0$. The only difference with systems with a lower
percentage of inhibitory neurons ($\chi_E=70\%,\chi_I=30\%$), is that
positive $\Delta_0$ are also observed in the fourth quadrant for small
values of $\delta EI$. This behavior can be attributed to the
different size of the two populations, since for the same $\delta$
values the inhibitory activity is not sufficient to balance the
excitatory one. Data confirm that for very small $\delta IE$ the
system is always imbalanced in favor of excitation (supercritical
behavior), whereas inhibition slightly overcomes excitation
(subcritical behavior) in the first quadrant parameter region.
Finally, we observe that, as the deterministic solutions from Eq.~(\ref{determin22}) for $E_0$ and $I_0$ are independent of the population
size, the value of $\Delta_0$ for ($\chi_E=\chi_I=50\%$) should be always smaller than that for ($\chi_E=70\%,\chi_I=30\%$).

\subsection{Eigenvalues}
\label{eigen}

Next, we calculate the eigenvalues for different values of $\delta EI$
and $\delta IE$. We stress that, under the hypothesis that synaptic
connections solely depend on the presynaptic neuron, the matrix ${\bf
A}$ always has real eigenvalues, corresponding to the two inverse
characteristic times in the correlation functions. In the present,
more general case the eigenvalues can become complex. In
Fig.\ref{eigen1}(top left) we show in different colors the parameter regions
where eigenvalues are real (cyan-colored region) and complex
(red-colored region). The eigenvalues are independent of the size of
the excitatory or inhibitory populations (see Section~\ref{sec1}) and have a
non-zero imaginary part in two regions of the parameter space. In the
diagonal region along the $\delta EI=-\delta IE$ line the imaginary
part of the eigenvalues is very small, close to zero. Conversely, in
the horizontal region the imaginary part can assume a wide range of
values, mainly depending on $\delta IE$ (Fig.\ref{eigen1} top right). 
 Fig.\ref{eigen1}(bottom) shows the  real part of the
eigenvalues $\lambda_1$ and $\lambda_2$. The eigenvalues are all negative, 
indicating that the system is stable for any value of the parameters inside this
region of the parameter space:
we computed the long time limit of Eq.(\ref{wc_deterministic}), that therefore brings the system to an attractive fixed point. In the first quadrant, corresponding to $\Sigma_0\simeq 0$ and $\Delta_0\simeq 0$, the real part of the eigenvalue $\lambda_1$ becomes constant and close to zero  whereas $\lambda_2$ exhibits large negative values. In the rest of the parameter space, whereas $\lambda_1$ assumes almost constant values, $\lambda_2$ appears to depend solely on the parameter $\delta {IE}$, becoming more negative for decreasing synaptic strengths. We will now
evaluate the correlation functions and the corresponding response
functions at different locations of the parameter space (points A to G).

\begin{figure}[ht]
        \centering
        \subfigure{
                \includegraphics[width=0.35\textwidth]{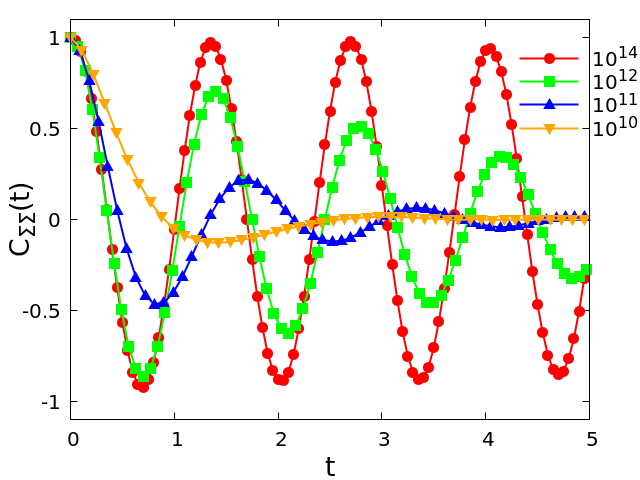}}

        \caption{The correlation function $C_{\Sigma\Sigma}(t)$ for $\chi_E=\chi_I=50\%$ evaluated at point C in the phase diagram Fig.\ref{eigen1},
          for different number of neurons $N=10^{10}$, $10^{11}$, $10^{12}$, $10^{14}$, with parameters  $w_{EE}=6.95$, $w_{II}=6.85$, $h=10^{-6}$, $\alpha=0.1$, $w_{EI}=w_{II}+\delta {EI}$ and $w_{IE}=w_{EE}+\delta{IE}$. Details of simulations are reported in Sec.~\ref{sec_numeric}.}
        \label{nfinito}
\end{figure}

\begin{figure*}[ht]
 	\centering
 	\subfigure{
 		\includegraphics[width=0.35\textwidth]{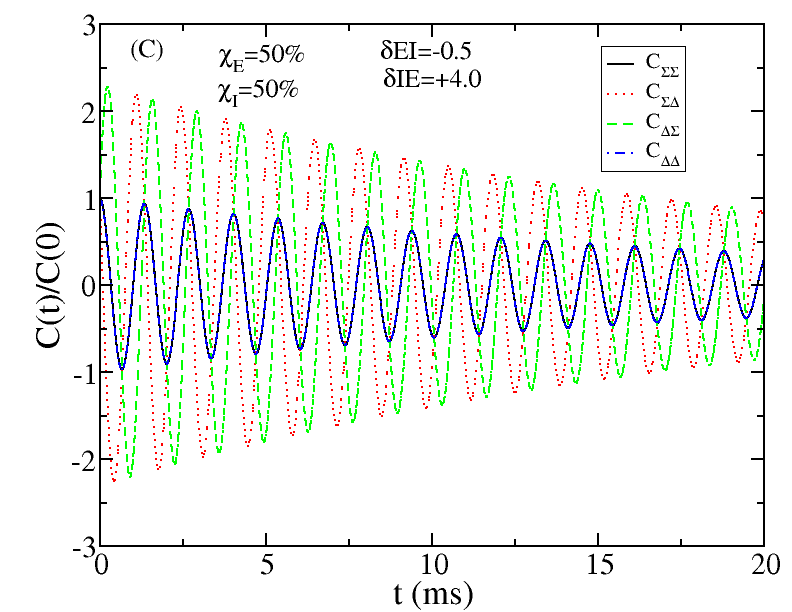}}
 	\subfigure{
 		\includegraphics[width=0.35\textwidth]{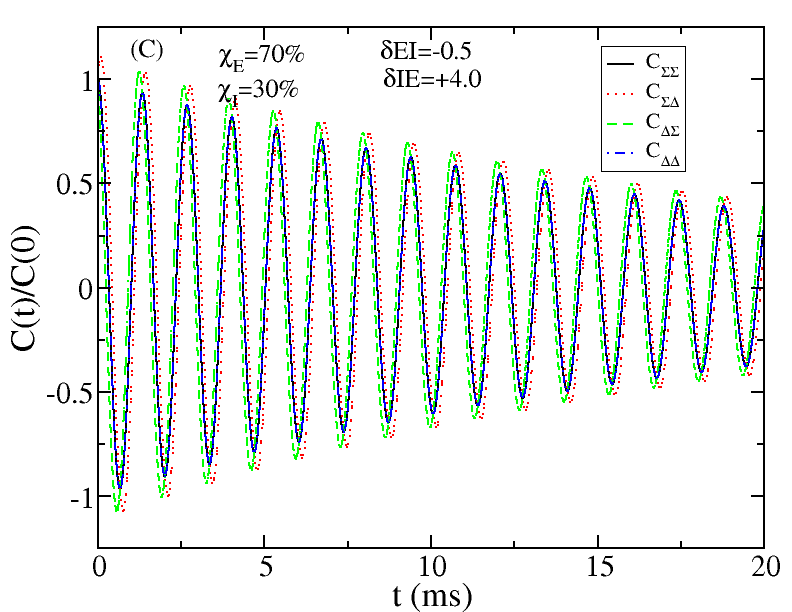}}
 	\subfigure{
 		\includegraphics[width=0.35\textwidth]{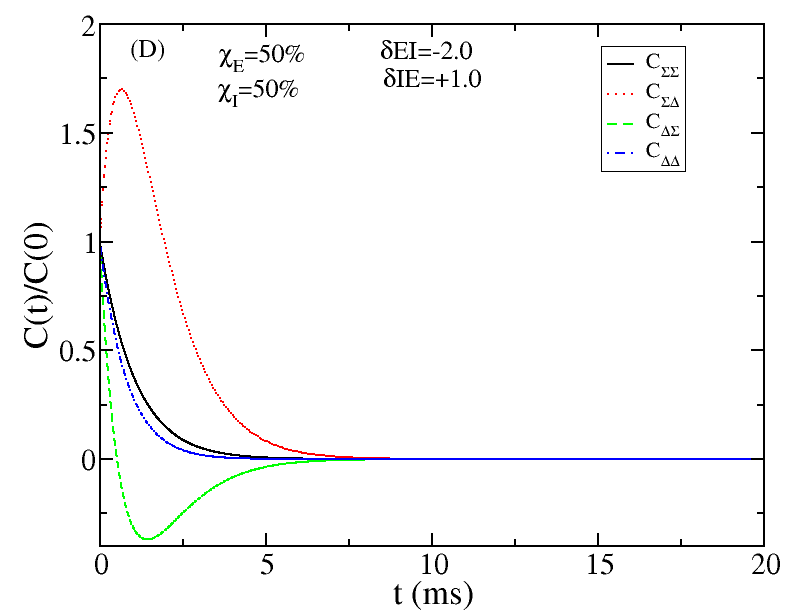}}
 	\subfigure{
 		\includegraphics[width=0.35\textwidth]{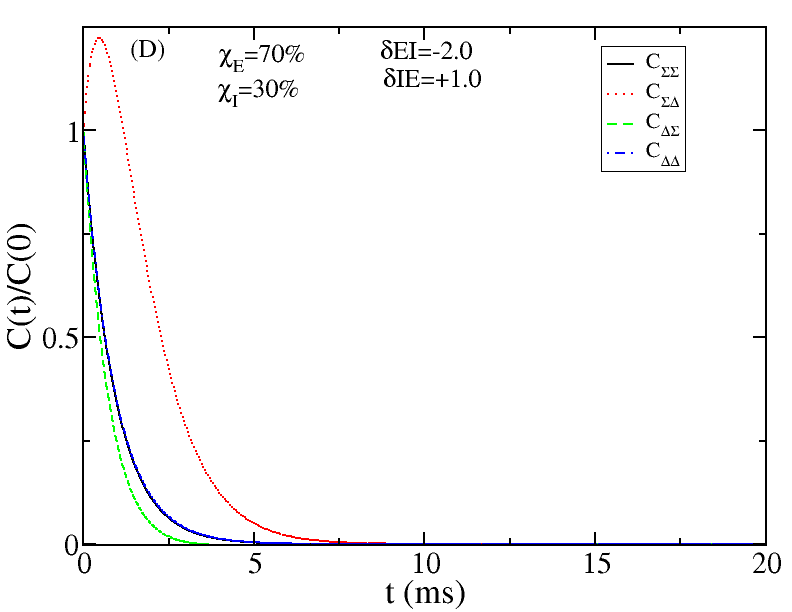}}
 	\centering
 	\subfigure{
 		\includegraphics[width=0.35\textwidth]{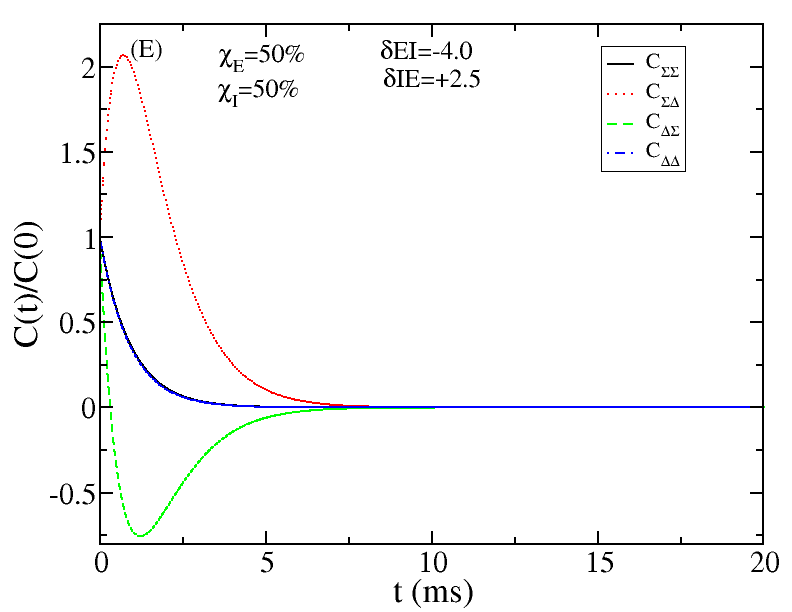}}
 	\subfigure{
 		\includegraphics[width=0.35\textwidth]{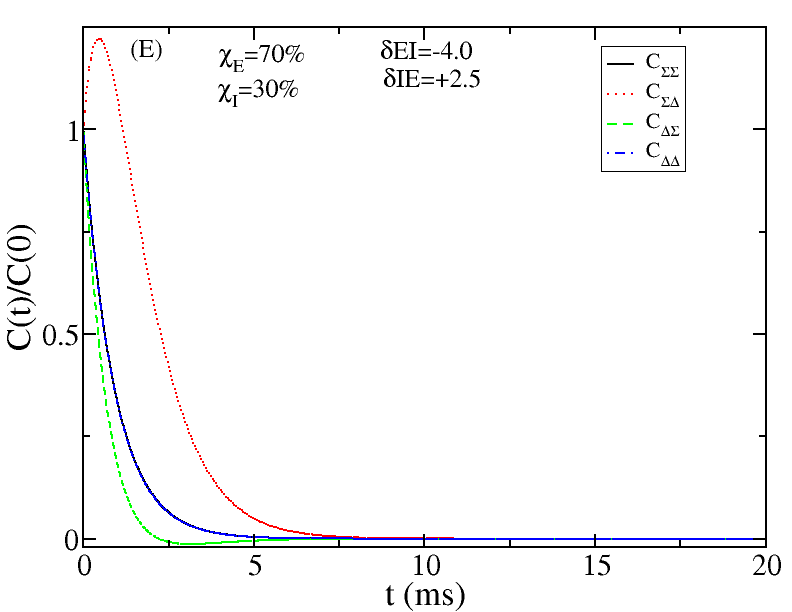}}
 	\subfigure{
 		\includegraphics[width=0.35\textwidth]{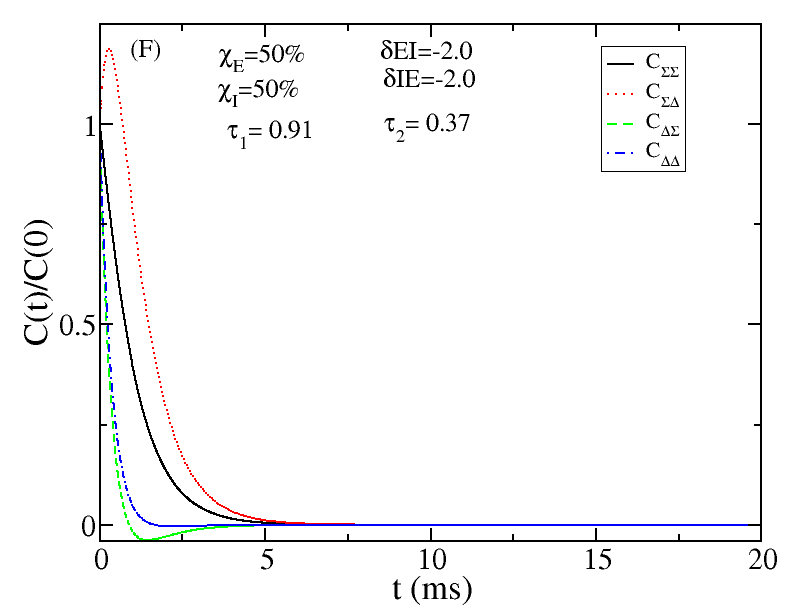}}
 	\subfigure{
 		\includegraphics[width=0.35\textwidth]{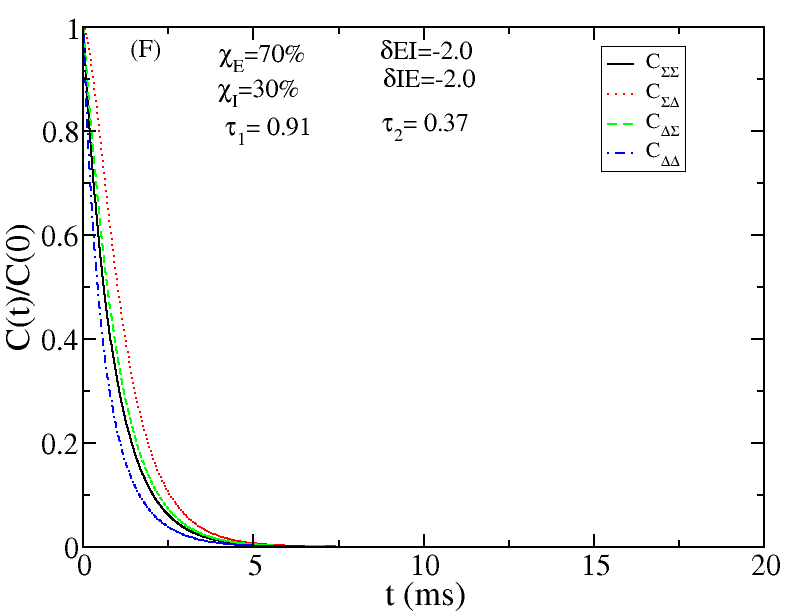}}
 	\caption{The auto- and cross-correlation functions at the different points  in the parameter space Fig.\ref{eigen1}.
The left column data are for systems with $\chi_E=\chi_I=50\%$, whereas the right column is for $\chi_E=70\%$ and $\chi_I=30\%$. Parameters are $w_{EE}=6.95$, $w_{II}=6.85$, $h=10^{-6}$, $\alpha=0.1$, $w_{EI}=w_{II}+\delta {EI}$ and $w_{IE}=w_{EE}+\delta{IE}$.}
 	\label{corr_analytic_all}
 \end{figure*}

\begin{figure*}[ht]
	\centering
	\subfigure{
		\includegraphics[width=0.35\textwidth]{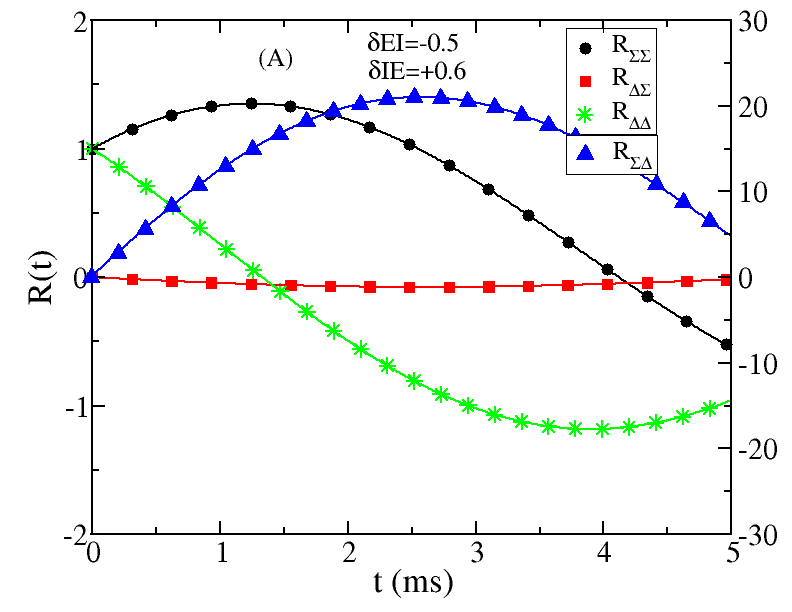}}
	\subfigure{
		\includegraphics[width=0.35\textwidth]{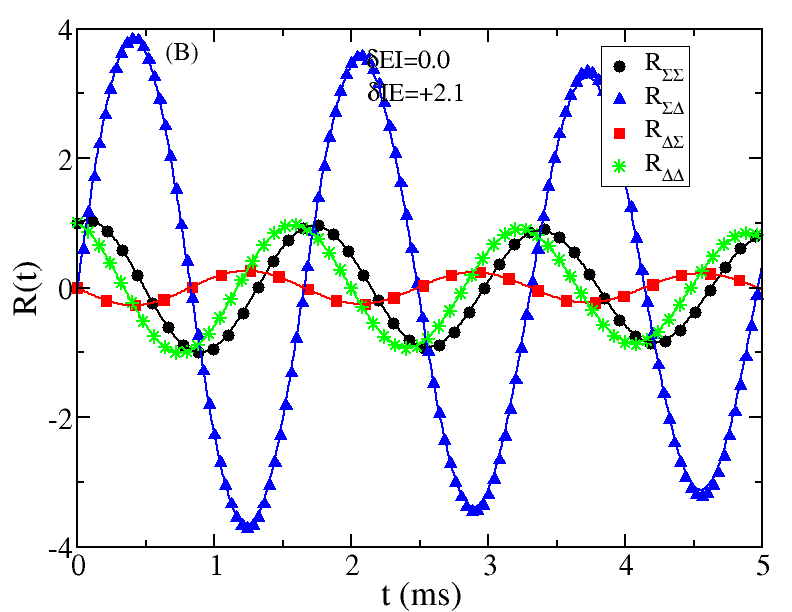}}
	\subfigure{
		\includegraphics[width=0.35\textwidth]{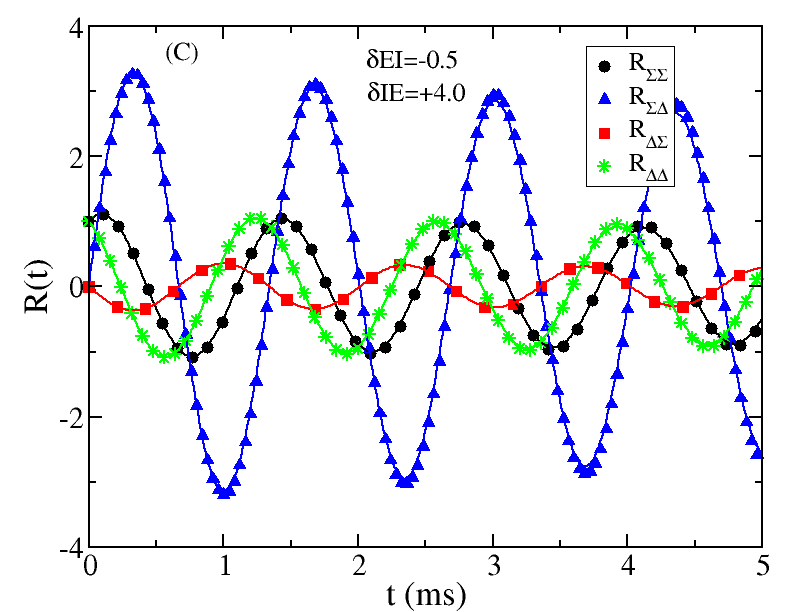}}
	\subfigure{
		\includegraphics[width=0.35\textwidth]{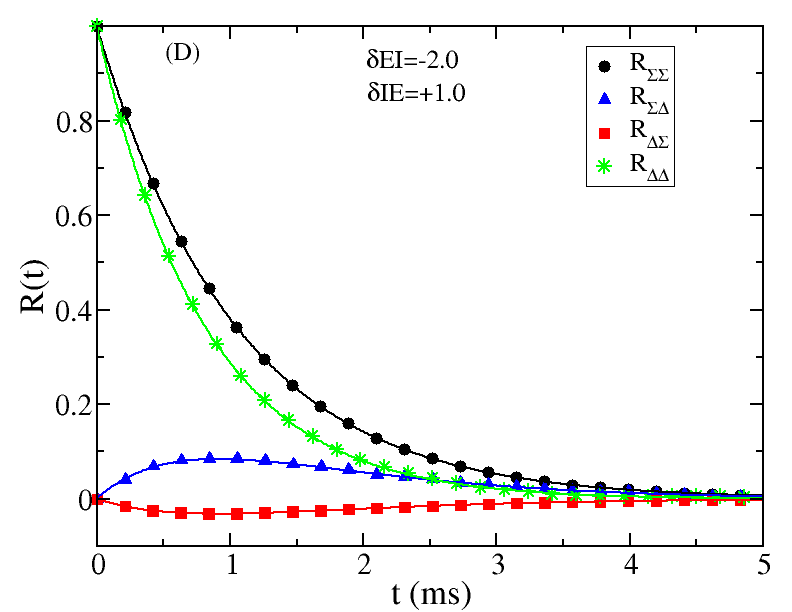}}
	\subfigure{
		\includegraphics[width=0.35\textwidth]{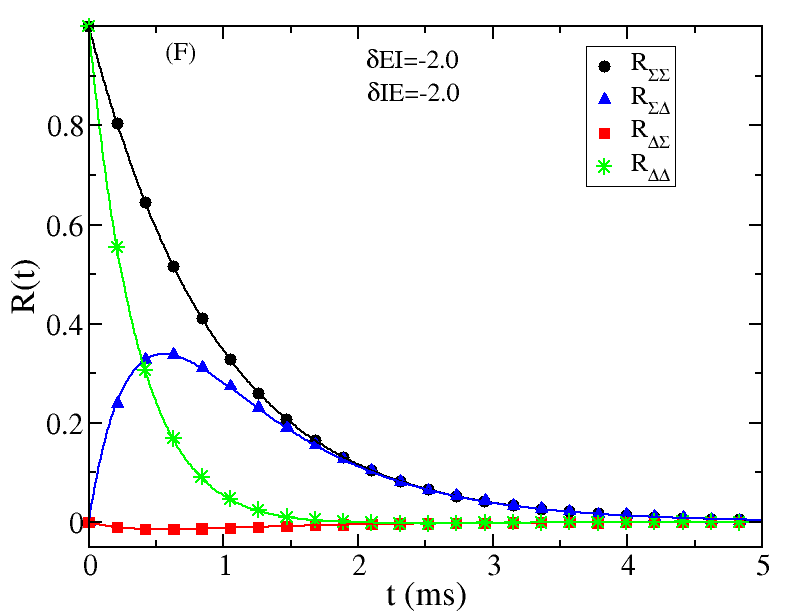}}
	\subfigure{
		\includegraphics[width=0.35\textwidth]{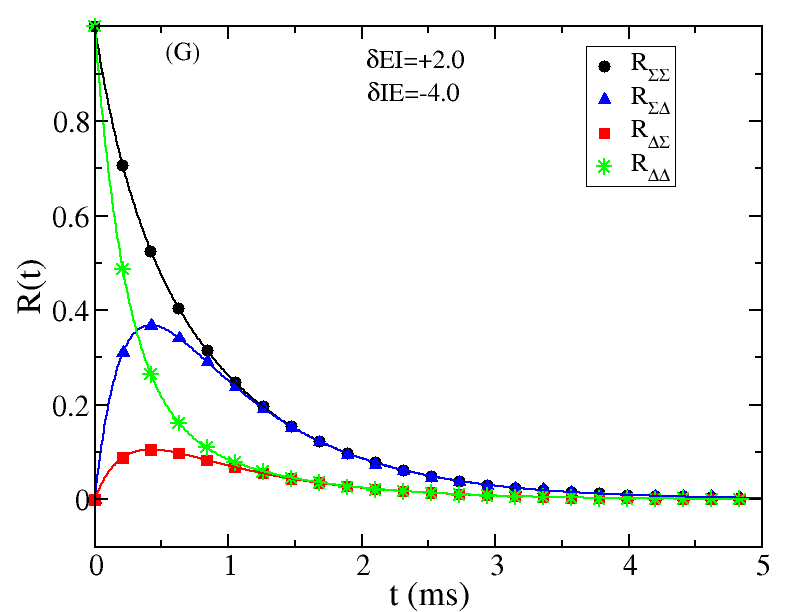}}
	\caption{The response functions for $\chi_E=\chi_I=50\%$ evaluated at different points A, B, C, D, F and G in the phase diagram Fig. \ref{eigen1}. The symbols are for numerical data ($N=10^{14}$),  with parameters  $w_{EE}=6.95$, $w_{II}=6.85$, $h=10^{-6}$, $\alpha=0.1$, $w_{EI}=w_{II}+\delta {EI}$ and $w_{IE}=w_{EE}+\delta{IE}$, whereas solid lines represent analytical results. The range of values for $R_{\Sigma\Delta}$ is [-30:30] only for point A. Details of simulations are reported in Sec.~\ref{sec_numeric}.}
\label{response_fig1}
\end{figure*}

\section{Correlation Functions}
\label{corr}

We analyse next the auto-correlation and cross-correlation functions at the points reported in the parameter space (Fig.\ref{eigen1}) by analytical calculation of Eqs.(\ref{cor11}-\ref{im_cor}). 
The correlation functions show (Fig.\ref{corr_func_analytic}) either a double exponential decay or oscillations depending on whether the eigenvalues are real or complex. Indeed, oscillations are observed at points B and C, with a frequency given by the absolute value of the imaginary part of the eigenvalue and independent of the population size. Interestingly, the frequency, evaluated for different $\delta EI$,
scales with  $\delta IE$ with an exponent smaller than one (see inset of Fig.\ref{corr_func_analytic}), confirming the important role of the activity of the inhibitory population in the system dynamics. Moreover, we observe that for systems with equal populations ($\chi_E=\chi_I=50\%$), there are strong phase differences and large amplitude differences between the self and cross-correlation functions, which for the system with $\chi_E=70\%$, $\chi_I=30\%$ (values typical of mammalian brains) is almost absent (see points A and B). On the other hand, for real eigenvalues, the analytical solution provides correlation functions which are a double exponential with characteristic times which do not depend on the population size. For instance, the fitting procedure gives $\tau_1=0.92$ and $\tau_2=0.23$ at point G. Interestingly, in the first quadrant the exponential containing the eigenvalue $\lambda_1$ has an amplitude close to zero ($\simeq 10^{-8}$), therefore the correlation functions all exhibit a sharp single exponential decay.

Next, we compare our analytical data with the data obtained from simulations of the Wilson-Cowan model using the Gillespie algorithm, or the non-linear Langevin equations
when the number of neurons is large.
Fig.\ref{corr_compare} shows the plots of the auto- and cross-correlation functions obtained from numerical simulations and analytical predictions
for equal population systems at all the points shown in the parameter space in Fig. \ref{eigen1}.
We find that the analytical predictions are perfectly matching with the simulated data, provided that the number of neurons used in simulations is large enough.
Namely, simulations must be performed with a total number of neurons $N>\overline\Sigma^{-2}$,
where $\overline\Sigma$ is the mean value of the activity.
Indeed, fluctuations in the activity are of order $N^{1/2}$, and the linear approximation Eq.(\ref{stocastic_sigma1}) is valid only when
fluctuations are smaller than the mean values.
This is particularly relevant at points in phase space where the mean value $\overline\Sigma$ is very low, for example at point C of Fig.\ref{eigen1} where
the mean value is $\overline\Sigma\sim 10^{-7}$.
In Fig.\ref{nfinito} we show the autocorrelation $C_{\Sigma\Sigma}(t)$ at point C for different number of neurons $N$.
We see a strong dependence of the results on $N$ up to $N\sim 10^{14}$.

The correlation functions evaluated at all the different points marked
in the parameter space are shown in
Fig. \ref{corr_analytic_all}. One observes high
  frequency oscillations for $\delta EI=-0.5$ and $\delta IE=+4$
  (point C), for all correlation functions, for both the considered
  fractions of exitatory and inhibitory populations. The other points
  D, E and F show a simple exponential decay for the auto-correlations
  of $\Sigma$ and $\Delta$, while cross-correlations are characterized
  by nonmonotonic behavior, featuring also a negative region for
  $C_{\Delta\Sigma}$ in the case of equal fraction populations. This
  phenomenon, pronounced in cases D and E, can be intepreted as a
  ``backscattering'' in $\Delta$ activity, namely a negative
  fluctuation of $\Delta$ that follows a positive fluctuation of
  $\Sigma$ after a certain time.

\section{Response Functions}
\label{resp}

We next calculate the response of the system to small instantaneous
perturbations according to
Eq. (\ref{response_eq3}). Fig. \ref{response_fig1} shows the four
different response functions for equal population systems. Under the
hypothesis of synaptic strengths depending solely on the presynaptic
neuron type, previous calculations \cite{sarracino2020predicting} have
shown that, due to the upper triangular form of the coupling matrix
$\bf A$, the response function exhibits a simple exponential decay
behavior ($R_{\Sigma\Sigma}$ and $R_{\Delta\Delta}$), a double
exponential decay ($R_{\Sigma\Delta}$) or vanishes
($R_{\Delta\Sigma}$). The single exponential stems for the fact that
the cross-correlation term cancels out one exponential decay. In the
imbalanced case, the response functions show a more complex behavior,
with oscillations at the points in parameter space where eigenvalues
are complex.

To obtain the response function from simulations,
and therefore compare analytical prediction to numerical data, 
we apply a weak perturbation to the system, in order to remain in the linear regime.
We take an equilibrium configuration, namely a configuration at
stationarity, we increase the value of $\xi_\Sigma$ or $\xi_\Delta$ by a small amount $\epsilon$
and compute $\epsilon^{-1}\langle\xi_\Sigma(t)\rangle$ or $\epsilon^{-1}\langle\xi_\Delta(t)\rangle$, respectively, at subsequent times,
as described in the Section \ref{sec_numeric}.
Fig.\ref{response_fig1} shows the comparison of the response functions between analytical calculation and simulation
data for all the points shown in the eigenvalue phase diagram Fig.\ref{eigen1}.
The symbols represent the simulation data, whereas the solid lines
 the data obtained from the analytical calculations of the response functions using Eq. (\ref{response_eq3}).
As in the case of correlations, numerical data and analytical calculations match well provided the number of neurons
is large enough.  Moreover, in this case a small value of $\epsilon$ has to be chosen:
data in Fig.\ref{response_fig1} were obtained with $\epsilon=10^{-3}$.

\section{Conclusions}
\label{con}
Each neuron in the brain can receive thousands of excitatory and inhibitory synaptic inputs. In physiological conditions, the ratio of
excitatory to inhibitory inputs remains stable at both single cell and global circuit levels, a property named balance of excitation and inhibition (EI) \cite{,zhou, isaacson}. Although
the existence of EI balance in the mammalian cortex has been widely studied and its disruption has been implicated in many brain diseases affecting higher cognitive functions, it is not yet clear how this balance is maintained in healthy brains \cite{bathia, hecline}.
Experimentally, imbalance arises hampering excitatory or inhibitory neurotransmission with selected antagonists \cite{beggs2003}. In neuronal networks, imbalance is obtained controlling the percentage of inhibitory synapses, the connectivity network or the neuron excitability, and can lead to an excess of large bursts, as observed in epileptic systems.
Several experimental and theoretical studies have confirmed that imbalanced conditions alter spontaneous brain activity. Imbalance modifies the typical scale-free behavior of activity in the resting state \cite{beggs2003, massobrio} and also its temporal features \cite{lombardi, chaos}. Analogously, EI balance and imbalance may affect the relation between spontaneous and evoked activity, i.e. the response to external stimuli \cite{arieli1996dynamics}.
Recently, the problem has been addressed theoretically by means of the FDRs connecting the
spontaneous fluctuations of a system with the response function to external perturbations \cite{sarracino2020predicting}. The analytical derivation, based on
the linear noise approximation of the Wilson Cowan model, on the main assumption of EI balance provides a double exponential decay for the correlation functions and a simple exponential for the $R_{\Sigma\Sigma}$. In this  study, we investigated its extension to imbalanced conditions in a wide range of parameters tuning such imbalance.

Results indicate that the main parameter controlling activity in imbalance is $\delta IE$, namely the variation in the synaptic strength exciting the inhibitory population. Conversely, the other parameter $\delta EI$, expressing the strength of inhibition received by the excitatory neurons, appears to have a different role: It controls the transition from a high activity regime to a regime (in the first quadrant) where $\Sigma_0\sim \Delta_0\sim0$, as well as the transition from real to complex eigenvalues.
The overall behavior of the system stems from the interplay between these two independent effects. The presence of complex eigenvalues leads to
a novel oscillatory behavior for the correlation functions in a narrow range of $\delta EI$ and, consequently, oscillations in the response functions, with a frequency depending on the parameter $\delta IE$. The important remark is that analytical results are fully confirmed by Gillespie simulations of Wilson Cowan networks in the limit of very large system size. Indeed, this limit, implemented to derive the FDRs, results to be extremely stringent since full agreement with simulation data is achieved for systems as large as $N\sim10^{14}$ neurons. Interestingly, the FDRs are fulfilled numerically even for smaller system sizes, where the agreement with the analytical solution is not perfect. The present results, obtained for a population model, are also in good agreement with simulations of integrate and fire networks models \cite{Raimo2020RoleOI}, where oscillations in the correlation functions were observed in the supercritical regime and the frequency depended on the percentage of inhibitory neurons and their level of connectivity.

\begin{acknowledgments}
	LdA and ADC would like to thank MIUR project PRIN2017WZFTZP
        for financial support. AS acknowledges support from MIUR
        project PRIN201798CZLJ. HJH thanks the University of Campania
        for the visiting professorship and FUNCAP for financial
        support. Work supported by \# NEXTGENERATIONEU (NGEU) and funded
        by the Ministry of University and Research (MUR), National
        Recovery and Resilience Plan (NRRP), project MNESYS
        (PE0000006) - A Multiscale integrated approach to the study
        of the nervous system in health and disease (DN. 1553
        11.10.2022).
\end{acknowledgments}
\section{Appendix}

\subsection{Stochastic Wilson-Cowan model}
We present here the derivation of the Wilson-Cowan equations in the general case where no assumption is made on the synaptic strengths and the neuronal populations can have different sizes. The dynamics evolves according to a master equation for the probability $p_{k,l} (t)$, where the number of active excitatory and active inhibitory neurons are $k$ and $l$. We set
\begin{eqnarray}
 k&=&N_EE+N_E^{1/2}\xi_E\nonumber\\
 l&=&N_II+N_I^{1/2}\xi_I
\end{eqnarray}
The input currents are
\begin{eqnarray}
 S_E&=&\frac{w_{EE} k}{N_E}-\frac{w_{EI} l}{N_I}+h_E\nonumber\\
 S_I&=&\frac{w_{IE} k}{N_E}-\frac{w_{II} l}{N_I}+h_I
\end{eqnarray}
The master equation, describing the evolution of the
probabilities $p_{k,l}(t)$ that the system is in the state $(k,l)$ at time $t$, is
\begin{eqnarray}
 \frac{dp_{k,l}(t)}{dt}=&&\alpha [(k+1)p_{k+1,l}(t)-kp_{k,l}(t)]\nonumber\\
 &&+[(N_E-k+1)f(S_E(k-1,l))p_{k-1,l}(t)\nonumber\\
 &&-(N_E-k)f(S_E(k,l))p_{k,l}(t)]\nonumber\\
 &&+\alpha [(l+1)p_{k,l+1}(t)-lp_{k,l}(t)]\nonumber\\
 &&+[(N_I-l+1)f(S_I(k,l-1))p_{k,l-1}(t)\nonumber\\
 &&-(N_I-l)f(S_I(k,l))p_{k,l}(t)]
\end{eqnarray}
Now using $e^{\partial_k}\{kp_{k,l}(t)\}=(k+1)p_{k+1,l}(t)$, we have

\begin{eqnarray}
\frac{dp_{k,l}(t)}{dt}=&&\alpha(e^{\partial_k}-1)kp_{k,l}(t)+(e^{-\partial_k}-1)\nonumber\\
&&[(N_E-k)f(S_E(k,l))p_{k,l}(t)]+\alpha(e^{\partial_l}-1)lp_{k,l}(t)\nonumber\\
&&+(e^{-\partial_l}-1)[(N_I-l)f(S_I(k,l))p_{k,l}(t).
\end{eqnarray}
Finally we have
\begin{eqnarray}
 \frac{dp_{k,l}(t)}{dt}=&&-\partial_k[N\chi_EA_E\Big(\frac{k}{N_E},\frac{l}{N_I}\Big)p_{k,l}(t)]\nonumber\\
 &&+\frac{1}{2}\partial_k^2[N\chi_ED_E\Big(\frac{k}{N_E},\frac{l}{N_I}\Big)p_{k,l}(t)]\nonumber\\
 &&-\frac{1}{3!}\partial_k^3[N\chi_EA_E\Big(\frac{k}{N_E},\frac{l}{N_I}\Big)p_{k,l}(t)]+...\nonumber\\
 &&-\partial_l[N\chi_IA_I\Big(\frac{k}{N_E},\frac{l}{N_I}\Big)p_{k,l}(t)]\nonumber\\
 &&+\frac{1}{2}\partial_l^2[N\chi_ID_I\Big(\frac{k}{N_E},\frac{l}{N_I}\Big)p_{k,l}(t)]\nonumber\\
 &&-\frac{1}{3!}\partial_l^3[N\chi_IA_I\Big(\frac{k}{N_E},\frac{l}{N_I}\Big)p_{k,l}(t)]+...
 \label{master0}
\end{eqnarray}
where
\begin{eqnarray}
 A_E(x,y)=&&-\alpha x+(1-x)f(w_{EE}x-w_{EI}y+h_E)\\
 D_E(x,y)=&&\alpha x+(1-x)f(w_{EE}x-w_{EI}y+h_E)\\
 A_I(x,y)=&&-\alpha y+(1-y)f(w_{IE}x-w_{II}y+h_I)\\
 D_I(x,y)=&&\alpha y+(1-y)f(w_{IE}x-w_{II}y+h_I),
\end{eqnarray}
with $x=\frac{k}{N_E}$, $y=\frac{l}{N_I}$, $\chi_E=\frac{N_E}{N}$ and $\chi_I=\frac{N_I}{N}$.

Now for $p_{k,l}(t)=\pi(\xi_E,\xi_I,t)$, we can write
\begin{eqnarray}
 \frac{\partial p_{k,l}(t)}{\partial t}=\frac{\partial}{\partial t}\pi(\xi_E,\xi_I,t)&&+\frac{\partial}{\partial t}\xi_E\frac{\partial}{\partial \xi_E}\pi(\xi_E,\xi_I,t)\nonumber\\
 &&+\frac{\partial}{\partial t}\xi_I\frac{\partial}{\partial \xi_I}\pi(\xi_E,\xi_I,t).
 \label{p2pi}
\end{eqnarray}
Considering that $\partial_t\xi_E=-N_E^{1/2}\partial_tE$ and $\partial_t\xi_I=-N_I^{1/2}\partial_tI$, from Eq.(\ref{p2pi}) we can write
\begin{eqnarray}
\partial_tp_{k,l}(t)= \partial_t\pi(\xi_E,\xi_I,t)&&-N_E^{1/2}\partial_tE\partial \xi_E\pi(\xi_E,\xi_I,t)\nonumber\\
&&-N_I^{1/2}\partial_tI\partial\xi_I\pi(\xi_E,\xi_I,t)
\label{p2pi2}
\end{eqnarray}
Next, by Taylor's expanding  $A_{E,I}$ and $D_{E,I}$ in powers of the system size, the leading term of the order $N^{1/2}$ provides the deterministic equations
\begin{eqnarray}
 -N^{1/2}\chi_E^{1/2}\partial_tE\partial_{\xi_E}\pi=-N^{1/2}\chi_E^{1/2}A_E(E,I)\partial_{\xi_E}\pi\nonumber\\
 -N^{1/2}\chi_I^{1/2}\partial_tI\partial_{\xi_I}\pi=-N^{1/2}\chi_I^{1/2}A_I(E,I)\partial_{\xi_I}\pi
 \label{determ1}
\end{eqnarray}
\begin{eqnarray}
 \frac{dE}{dt}=-\alpha E+(1-E) f(S_{E})\nonumber\\
  \frac{dI}{dt}=-\alpha I+(1-I) f(S_{I})
  \label{determin2}
\end{eqnarray}
whereas the successive term of the order $N^{0}$ provides the Fokker-Planck equations
\begin{eqnarray}
 \partial_t\pi=-[A_{E,E}(E,I)\partial_{\xi_E}(\xi_E\pi)+\sqrt{\chi_E/\chi_I}A_{E,I}(E,I) \nonumber\\
 \partial_{\xi_E}(\xi_I\pi)+\sqrt{\chi_I/\chi_E}A_{I,E}(E,I)\partial_{\xi_I}(\xi_I\pi) +A_{I,I}(E,I)\nonumber\\
\partial_{\xi_I}(\xi_I\pi)]+\frac{1}{2}D_E(E,I)\partial^2_{\xi_E}\pi+\frac{1}{2}D_I(E,I)\partial^2_{\xi_I}\pi
\end{eqnarray}
This approximation, which drops all successive terms, is named "linear noise approximation" and can be rewritten as two coupled Langevin equations
\begin{equation}
 \frac{d}{dt}\begin{pmatrix}
  \xi_E \\
  \xi_I
\end{pmatrix}={\bf \tilde{A}}\begin{pmatrix}
  \xi_E \\
  \xi_I
\end{pmatrix}+{\bf \tilde{D}}\begin{pmatrix}
  \eta_E \\
  \eta_I
\end{pmatrix}
\label{stocastic}
\end{equation}
where
${\bf \tilde{A}}=\begin{pmatrix}
  A_{E,E}(E,I) & \sqrt{\chi_E/\chi_I}A_{E,I}(E,I)\\
  \sqrt{\chi_I/\chi_E}A_{I,E}(E,I) & A_{I,I}(E,I)
\end{pmatrix}$
with
\begin{eqnarray}
 A_{E,E}(E,I)=-\alpha-f(S_E)+(1-E)w_{EE}f'(S_E)\\
 A_{E,I}(E,I)=-(1-E)w_{EI}f'(S_E)\\
 A_{I,E}(E,I)=(1-I)w_{IE}f'(S_I)\\
 A_{I,I}(E,I)=-\alpha-f(S_I)-(1-I)w_{II}f'(S_I)\\
 D_E(E,I)=\sqrt{\alpha E+(1-E)f(S_E)}\\
 D_I(E,I)=\sqrt{\alpha I+(1-I)f(S_I)}
\end{eqnarray}
and,
${\bf \tilde{D}}=\begin{pmatrix}
  D_E&0\\0&D_I
\end{pmatrix}$, 
where at the fixed point, $D_E=\sqrt{2\alpha E_0}$ and $D_I=\sqrt{2\alpha I_0}$.

By introducing the variable change from $(E,I)$ to $(\Sigma,\Delta)$, it is possible to obtain the set of deterministic equations (\ref{wc_deterministic}) and
from Eq.(\ref{stocastic}) the two coupled Langevin equations for the fluctuating terms
\begin{equation}
 \frac{d}{dt}\begin{pmatrix}
  \xi_\Sigma \\
  \xi_\Delta
\end{pmatrix}={\bf {A}}\begin{pmatrix}
  \xi_\Sigma \\
  \xi_\Delta
\end{pmatrix}+{\bf D}\begin{pmatrix}
  \eta_\Sigma \\
  \eta_\Delta
\end{pmatrix}
\label{stocastic_sigma_ap}
\end{equation}
where
\begin{eqnarray}
\begin{pmatrix}
  \xi_\Sigma \\
  \xi_\Delta
\end{pmatrix}= \begin{pmatrix}
 \chi_E & \chi_I\\ \chi_E&-\chi_I\end{pmatrix}\begin{pmatrix}
  \xi_E \\
  \xi_I
\end{pmatrix}
\end{eqnarray}

\begin{widetext}
\begin{eqnarray}
 {\bf A}= \begin{pmatrix}
 \chi_E & \chi_I\\ \chi_E&-\chi_I\end{pmatrix}\begin{pmatrix}
  A_{E,E}(E,I) & \sqrt{\chi_E/\chi_I}A_{E,I}(E,I) \\
  \sqrt{\chi_I/\chi_E}A_{I,E}(E,I)& A_{I,I}(E,I)
\end{pmatrix}\frac{1}{2}\begin{pmatrix}
\frac{1}{\chi_E} &\frac{1}{\chi_E}\\
\frac{1}{\chi_I} & -\frac{1}{\chi_I}
  \end{pmatrix}=\begin{pmatrix}
    x&y\\z&w
  \end{pmatrix}
\end{eqnarray}
\end{widetext}
with $x=\frac{1}{2}[A_{E,E}(E,I)+\Big(\frac{\chi_E}{\chi_I}\Big)^{3/2}A_{E,I}(E,I)+\Big(\frac{\chi_I}{\chi_E}\Big)^{3/2}A_{I,E}(E,I)+A_{I,I}(E,I)]$,
      $y=\frac{1}{2}[A_{E,E}(E,I)-\Big(\frac{\chi_E}{\chi_I}\Big)^{3/2}A_{E,I}(E,I)+\Big(\frac{\chi_I}{\chi_E}\Big)^{3/2}A_{I,E}(E,I)-A_{I,I}(E,I)]$,
      $z=\frac{1}{2}[A_{E,E}(E,I)+\Big(\frac{\chi_E}{\chi_I}\Big)^{3/2}A_{E,I}(E,I)-\Big(\frac{\chi_I}{\chi_E}\Big)^{3/2}A_{I,E}(E,I)-A_{I,I}(E,I)]$ and
      $w=\frac{1}{2}[A_{E,E}(E,I)-\Big(\frac{\chi_E}{\chi_I}\Big)^{3/2}A_{E,I}(E,I)-\Big(\frac{\chi_I}{\chi_E}\Big)^{3/2}A_{I,E}(E,I)+A_{I,I}(E,I)]$.
   Moreover,
    \begin{eqnarray}
     {\bf D}=&&\begin{pmatrix}
 \chi_E & \chi_I\\ \chi_E&-\chi_I\end{pmatrix}\begin{pmatrix}
  D_E & 0 \\
  0& D_I
\end{pmatrix}\frac{1}{2}\begin{pmatrix}
\frac{1}{\chi_E} &\frac{1}{\chi_E}\\
\frac{1}{\chi_I} & -\frac{1}{\chi_I}
  \end{pmatrix}\nonumber\\
  &&=\frac{1}{2}\begin{pmatrix}
    D_E+D_I&D_E-D_I\\D_E-D_I&D_E+D_I
  \end{pmatrix} 
    \end{eqnarray}
The noise amplitude matrix can be written as 
\begin{eqnarray}
 {\bf \mathcal{ M}}={\bf D D}^T=\begin{pmatrix}
                               G & H\\
                               H& G
                              \end{pmatrix},
\end{eqnarray}
where at the fixed points $G=\frac{1}{2}(D_E^2+D_I^2)=\alpha E_0+\alpha I_0$ and $H=\frac{1}{2}(D_E^2-D_I^2)=\alpha E_0-\alpha I_0$.
The Eq.(\ref{stocastic_sigma_ap}) can be written in more compact form as
\begin{eqnarray}
 \frac{d{\bf X}}{dt}={\bf AX}+{\bf D}{\bm \eta},
 \label{compact_eq_ap}
\end{eqnarray}
where ${\bf X}\equiv(\xi_\Sigma, \xi_\Delta)$ and ${\bm {\eta}}\equiv(\eta_\Sigma,\eta_\Delta)$. Then we can write the solutions of the above equation as
\begin{equation}
 {\bf X}(t)=e^{{\bf A}t}{\bf X}(0)+{\bf D}\int_0^t dt' e^{{\bf A}(t-t')}{\bm \eta}(t'),
\label{compact_soln_ap}
\end{equation}
valid for $t\geq 0$.


\subsection{Correlation functions}
The correlation matrix for the fluctuating terms is defined as
\begin{equation}
 C_{ij}(t)\equiv\langle\xi_i(t)\xi_j(0)\rangle=(e^{ At}\sigma)_{ij},
 \label{corr1_ap}
\end{equation}
where ${\sigma_{ij}}=\langle\xi_i(0)\xi_j(0)\rangle$ are the components of the covariance matrix which satisfies
\begin{eqnarray}
{\bf \mathcal{M}}=\frac{{\bf A}\sigma+\sigma {\bf A}^T}{2}.
\end{eqnarray}
The covariance matrix can be written as
\begin{eqnarray}
 \sigma=\begin{pmatrix}
  \sigma_{11} &\sigma_{12}\\
   \sigma_{21}&\sigma_{22}
 \end{pmatrix}
\end{eqnarray}
where
\begin{eqnarray}
 \sigma_{11}=\frac{G-\sigma_{12}y}{x}\\
 \sigma_{12}=\sigma_{21}=-\frac{G(zw+xy)-2Hxw}{(x+w)(xw-yz)}\\
 \sigma_{22}=\frac{G-\sigma_{12}z}{w}.
\end{eqnarray}\\

The eigenvalues of the matrix ${\bf A}$ are $\lambda_{\pm}=\frac{1}{2}[(x+w)\pm \sqrt{-\Theta}]$, where $\sqrt{-\Theta}=\sqrt{(x-w)^2+4 yz}$.\\
To evaluate $e^{{\bf A}t}$ we need a diagonalizing matrix ${\bf P}$, which can be written as
$$P=\begin{pmatrix}
       \frac{(x-\lambda_1)}{z\sqrt{1+[(x-\lambda_1)/z]^2}} & \frac{(x-\lambda_2)}{z\sqrt{1+[(x-\lambda_2)/z]^2}}\\
       \frac{1}{\sqrt{1+[(x-\lambda_1)/z]^2}} & \frac{1}{\sqrt{1+[(x-\lambda_2)/z]^2}}
       \end{pmatrix}.$$\\
       \\
Now let $\mu=\sqrt{1+[(x-\lambda_1)/z]^2}$ and $\nu=\sqrt{1+[(x-\lambda_2)/z]^2}$, then $\bf P$ becomes\\
$$P=\begin{pmatrix}
   \frac{x-\lambda_1}{z\mu} & \frac{x-\lambda_2}{z\nu}\\
   \frac{1}{\mu} & \frac{1}{\nu}
   \end{pmatrix}$$ whose determinant is 
   $|{\bf P}|=-\frac{\sqrt{-\Theta}}{z\mu\nu}$.
   
   The inverse matrix of $\bf P$ is 
\begin{eqnarray}
   P^{-1}=-\frac{z\mu\nu}{\sqrt{-\Theta}}\begin{pmatrix}
           \frac{1}{\nu}& -\frac{x-\lambda_2}{z\nu}\\
           -\frac{1}{\mu}& \frac{x-\lambda_1}{z\mu}
        \end{pmatrix}\nonumber\\
        =\frac{1}{\sqrt{-\Theta}}\begin{pmatrix}
           -{z\mu}& (x-\lambda_2)\mu\\
           {z\nu}& -(x-\lambda_1)\nu
        \end{pmatrix}.
\end{eqnarray}
Then, we can write the matrix exponential as
\begin{widetext}
\begin{eqnarray}
e^{At}&=&Pe^{\lambda_{\pm}t}P^{-1}\nonumber\\
&=&\frac{1}{\sqrt{-\Theta}}\begin{pmatrix}
\frac{x-\lambda_1}{z\mu} & \frac{x-\lambda_2}{z\nu}\\
   \frac{1}{\mu} & \frac{1}{\nu}
\end{pmatrix}\begin{pmatrix}
  e^{\lambda_2t}&0\\
  0&e^{\lambda_1t}
\end{pmatrix}\begin{pmatrix}
           -{z\mu}& (x-\lambda_2)\mu\\
           {z\nu}& -(x-\lambda_1)\nu
        \end{pmatrix}\nonumber\\
&=&\frac{1}{\sqrt{-\Theta}}\begin{pmatrix}
[(x-\lambda_2)e^{\lambda_1t}-(x-\lambda_1)e^{\lambda_2t}] & [y(e^{\lambda_1t}-e^{\lambda_2t})]\\
[z(e^{\lambda_1t}-e^{\lambda_2t})] &
[(x-\lambda_2)e^{\lambda_2t}-(x-\lambda_1)e^{\lambda_1t}]
                           \end{pmatrix}.
\label{expA}
\end{eqnarray}
\end{widetext}
Now from Eq.(\ref{corr1_ap}), calculating the matrix product, we
obtain the four correlation functions reported in the main text.

\subsection{Response functions}
The response of the system is defined as $ R_{ij}(t)\equiv \frac{\overline{\delta \xi_i(t)}}{\delta \xi_j(0)}$, with $(i,j)=(\Sigma, \Delta)$ as the response in $\xi_i$ once an instantaneous perturbation in $\xi_j$ is applied at $t=0$. According to Eq.(\ref{compact_soln_ap}), the response matrix is
\begin{equation}
 R(t)= e^{At}.
 \label{response_eq2_ap} 
\end{equation}
From Eq.(\ref{response_eq2_ap}), we see that the matrix exponential $e^{At}$ and the response function $R(t)$ coincide, therefore from Eq.(\ref{expA}) we can write the equations for the response functions as
\begin{eqnarray}
 R_{\Sigma\Sigma}(t)=&&\frac{1}{\sqrt{-\Theta}}[(x-\lambda_2)e^{\lambda_1t}-(x-\lambda_1)e^{\lambda_2t}]\nonumber\\
 R_{\Sigma\Delta}(t)=&&\frac{1}{\sqrt{-\Theta}}[y(e^{\lambda_1t}-e^{\lambda_2t})]\nonumber\\
 R_{\Delta\Sigma}(t)=&&\frac{1}{\sqrt{-\Theta}}[z(e^{\lambda_1t}-e^{\lambda_2t})]\nonumber\\
 R_{\Delta\Delta}(t)=&&\frac{1}{\sqrt{-\Theta}}[(x-\lambda_2)e^{\lambda_2t}-(x-\lambda_1)e^{\lambda_1t}].
 \label{rsponse_funcAp}
\end{eqnarray}
In the case of complex eigenvalues $\lambda_{\pm}=a\pm ib$, where $a=(x+w)/2$ and $ib=\sqrt{-\Theta}/2$, we can write the response functions as
\begin{eqnarray}
 R_{\Sigma\Sigma}(t)&=&e^{at}\cos(bt)+\frac{(x-a)}{b}e^{at}\sin(bt)\nonumber\\
R_{\Sigma\Delta}(t)&=&\frac{y}{b}e^{at}\sin(bt)\nonumber\\
R_{\Delta\Sigma}(t)&=&\frac{z}{b}e^{at}\sin(bt)\nonumber\\
R_{\Delta\Delta}(t)&=&e^{at}\cos(bt)-\frac{(x-a)}{b}e^{at}\sin(bt).\nonumber\\
\label{response_imaginary}
\end{eqnarray}

\bibliographystyle{h-physrev}
\bibliography{reference.bib}

\end{document}